\documentstyle[psfig]{article}
\begin{document}
\begin{center}
{\Large {\bf Analytical expressions for optimum alignment modes of
highly segmented mirrors}}\\
L. Noethe\\
European Southern Observatory, Karl-Schwarzschild-Stra\ss e 2,\\
D-85748 Garching bei M\"unchen, Germany
\end{center}
\begin{center}
{\bf Abstract}
\end{center}
\vspace{-5mm}
{\bf Abstract.} The major sources causing deterioration of optical quality in extremely
large optical telescopes are
misadjustments of the mirrors, deformations of monolithic mirrors, and
misalignments of segments in segmented mirrors. For active optics corrections,
all three errors, which can partially compensate each other, are measured simultaneously.
It is therefore of interest to understand the similarities and differences between the
three corresponding types of modes which describe these errors.
The first two types are best represented by Zernike polynomials
and elastic modes respectively, both of them being continuous and smooth functions. The
segment misaligment modes, which are derived by singular value decomposition,
are by their nature not smooth and in general discontinuous. However, for
mirrors with a large number of segments, the lowest modes
become effectively both smooth and continuous.
This paper derives analytical expressions for these modes, using differential
operators and their adjoints, for the limit case of infinitesimally small
segments.
For segmented mirrors with approximately 1000 segments, it is shown that these modes
agree well with the corresponding lowest singular value decomposition modes.
Furthermore, the analytical expressions reveal the nature of the segment misalignment
modes and allow for a detailed comparison with the elastic modes of monolithic mirrors.
Some mathematical features emerge as identical in the two cases.
\section{Introduction}
The first large optical telescope with a segmented mirror was the ten-meter Keck Telescope.
Its primary mirror consisted of 36 hexagonal segments.
All currently envisaged extremely large optical telescopes
will have segmented primary mirrors, but
with a much larger number of hexagonal segments, of the order of one thousand or more.
Assuming that the shapes of the segments are perfect at all times,
the only source for wavefront errors
generated by a segmented mirror would be misalignments of the segments.
These can be measured either optically or by edge sensors at intersegment edges.
Both types of measurements can, in principle, detect the relative piston and the relative tilt
of adjacent segments. The misalignments are then corrected by
appropriate vertical position changes of actuators which can modify piston,
tip and tilt of each segment [1].

If the signals given by the measurements were free of noise and the actuators were perfect,
the misalignments of the segments could be perfectly corrected. But, in reality, both
signal noise and correction noise will give rise to residual misalignments.
Both the signals and the actuator positions can be represented as vectors with suitable sets
of base vectors. If the errors are introduced by the actuators themselves, one can choose
an arbitrary orthogonal set of base vectors. However, if the errors are due to the signals,
an optimum modal correction requires the possibility to expand the signal into a set of
well distinguishable, that is orthogonal signal modes, which correspond to well
distinguishable, that is orthogonal, actuator modes.
Such modes are the optimum alignment modes. For mathematical reasons, also described in
section~\ref{sec:introMethod}, they were called normal modes in reference [1]. These modes
can be calculated by Singular Value Decomposition (SVD). In the rest of the paper we assume
that the actuators are perfect and noise is only introduced by the signals.

For segmented mirrors with a thousand segments or more the modes with the lowest so called
singular values
are rather smooth. They are also the ones which appear statistically
with the largest amplitudes
in an expansion of the misalignment error due to random signal noise. In modern large
active optical telescopes [2] containing both segmented and flexible monolithic
mirrors the lowest normal modes
could largely be compensated or corrected by other elements in the telescope optics.
This shows the need to compare the normal modes with
other sets of modes used in large telescopes, for example Zernike polynomials or elastic
modes of monolithic mirrors [3].

In the limit of an infinite number of segments over a finite area the lowest normal
modes should be continuous smooth functions. If the relationship between a position
function and the corresponding signal function is expressed by a continuous operator,
such smooth functions can be calculated from differential equations
and appropriate boundary conditions. It should therefore
be possible to derive analytical expressions for the lowest normal modes in the limit of
infinitesimally small segments.

Section~\ref{sec:introMethod} outlines the method to obtain the analytical solutions.
Before treating the case of two-dimensional segmentation the method is applied in
section~\ref{sec:1DimSolutions} to two examples of one-dimensional line segments.
In section~\ref{sec:analytical2Dim} the method is used to calculate analytical
expressions of the normal modes of a mirror with hexagonal segments and
section~\ref{sec:SVD2Dim} compares the analytical solutions with the SVD solutions
of a highly segmented mirror.
Section~\ref{sec:minRmsSignal} discusses the relationship to modes calculated
with variational methods to minimise the r.m.s. of the signal vector for a given
r.m.s. of the actuator vector and section~\ref{sec:compElastic} shows the relationship
to elastic minimum energy modes of a circular plate.
\section{Method to obtain analytical solutions}
\label{sec:introMethod}
A given set of actuator positions
is described by a vector ${\tilde {\bf v}}$
in a $n_{{\rm act}}$-dimensional vector space ${\cal A}$, and a given set of signals
by a vector ${\tilde {\bf u}}$ in a $n_{{\rm sig}}$-dimensional vector space ${\cal S}$.
Without noise effects, that is if the signals depend only on the actuator
positions, the actuator vectors are related to the signal vectors by a linear
mapping $A$ from ${\cal A}$ to ${\cal S}$, represented as a two-dimensional
$n_{{\rm sig}} \times n_{{\rm act}}$ matrix ${\bf A}$ :
\begin{equation}
  {\bf A} {\bf {\tilde v}} = {\bf {\tilde u}}
     \label {eq:connectionUVtilde}
\end{equation}
In this case a misalignment could be perfectly corrected by actuator changes.
In reality, noise in the signals is not negligible and only those
signal vectors which are in the image space of ${\cal A}$ under the mapping
$A$ can be fully corrected.
An optimum modal correction of a misalignment of the segments requires the possibility
to expand the signal in an orthogonal set $\{ {\bf u}_{i} \}$ of signal modes which
are related by the mapping $A$ to an orthogonal set $\{ {\bf v}_{i} \}$ of actuator modes.
Such sets are generated by the Singular Value Decomposition of ${\bf A}$ :
\begin{equation}
  {\bf A} = {\bf U}{\bf \Sigma} {\bf V}^{{\rm H}}
     \label {eq:svd}
\end{equation}
${\bf U}$ is a matrix of dimension $n_{{\rm sig}} \times n_{{\rm act}}$ and
${\bf \Sigma}$ and ${\bf V}^{{\rm H}}$, where the superscript H denotes the adjoint
matrix, are square matrices of dimension
$n_{{\rm act}} \times n_{{\rm act}}$. The matrix ${\bf \Sigma}$ is diagonal and contains
$n_{{\rm act}}$ so called singular values $\sigma_{{\rm svd},i}$.
The $n_{{\rm act}}$ column vectors of ${\bf U}$ and row vectors of
${\bf V}$ form orthonormal sets $\{ {\bf u}_{i} \}$ and $\{ {\bf v}_{j} \}$,
respectively, that is, with $<,>$ denoting the inner vector product,
\begin{equation}
   <{\bf u}_{i}, {\bf u}_{j}> = <{\bf v}_{i}, {\bf v}_{j}> = \delta_{i,j}
             \label{eq:orthoSetsUV}
\end{equation}
The vectors ${\bf u}_{i}$ and ${\bf v}_{i}$ which are related by
\begin{equation}
  {\bf A} {\bf v}_{i} = \sigma_{{\rm svd},i} \, {\bf u}_{i}
     \label {eq:connectionUV}
\end{equation}
will be called normal signal and actuator modes, respectively, since they are
eigenvectors of so called normal matrices,
which are matrices which commute with their adjoints.
The normal matrices are ${\bf A}^{{\rm H}}{\bf A}$ in the case of the actuator
vectors and ${\bf A}{\bf A}^{{\rm H}}$ in the case of the signal vectors.
In the following the term normal mode without a specification
'actuator' or 'signal' will refer to an actuator mode.

For a large number of $n_{{\rm seg}}$ of segments the normal modes with the smallest
singular values become rather smooth. In the limit of an infinite number of segments
over a finite area, that is in terms
of the radii $a$ of a segment (see figure~\ref{fig:figure5}) and $R$
of the mirror, $a/R \rightarrow 0$, they should converge to functions which
are continuously differentiable an infinite number of times.
Let $L$ denote the continuous operator which corresponds to the matrix ${\bf A}$,
and $f_{i}$ and $g_{i}$ the functions, called the analytical position
and signal modes, to which the actuator modes ${\bf v}_{i}$ and the signal modes
${\bf u}_{i}$, respectively, converge for $a/R \rightarrow 0$.
The position functions $f_{i}$ will also be referred to as analytical normal modes.
The two now infinite sets $\{ f_{i} \}$ and $\{ g_{i} \}$ will have, similar to the sets
$\{ {\bf v}_{i} \}$ and $\{ {\bf u}_{i} \}$, the following two characteristics.
Let the vector products be defined by
\begin{equation}
   <f_{i}, f_{j}> = \frac{1}{2D} \,
        \int_{-D}^{+D} {\rm d}x\, f_{i}(x) f_{j}(x)
     = \frac{1}{2} \,
        \int_{-1}^{+1} {\rm d}\xi\, f_{i}(\xi) f_{j}(\xi)
\end{equation}
for the one-dimensional problems with a length of the chain of segments of $2D$
and the normalised variable $\xi = x/D$ and by
\begin{equation}
   <f_{i}, f_{j}> = \frac{1}{\pi (R^{2} - r_{1}^{2})} \,
          \int_{0}^{2\pi} {\rm d}\varphi
          \int_{r_{1}}^{R} r\, {\rm d}r\, f_{i}(r,\varphi) f_{j}(r,\varphi)
     = \frac{1}{\pi (1 - \rho_{1}^{2})} \,
          \int_{0}^{2\pi} {\rm d}\varphi
          \int_{\rho_{1}}^{1} \rho\, {\rm d}\rho
            \, f_{i}(\rho,\varphi) f_{j}(\rho,\varphi)
\end{equation}
for the two-dimensional problems, where $r$ is the radial, $\varphi$ the azimuthal
coordinate, $R$ the outer radius of a circular mirror, $r_{1}$ the radius of its inner hole,
$\rho = r/R$ the normalised radial variable, and $\rho_{1} = r_{1}/R$ the normalised
radius of the inner hole. The first characteristic is then that
the two sets of functions are orthonormal, that is
\begin{equation}
  <f_{i}, f_{j}> = <g_{i}, g_{j}> = \delta_{i,j}
	\label{eq:orthoFG}
\end{equation}
The second characteristic is that the members $f_{i}$ and $g_{i}$ are related by
\begin{equation}
  L \; f_{i} = {\tilde \sigma}_{i} \, g_{i} \; , \label{eq:connectionFG}
\end{equation}
where the ${\tilde \sigma}_{i}$, which from now on will be called analytical singular values,
correspond to the singular values $\sigma_{{\rm svd},i}$ in the discrete SVD.

The functions $f_{i}$ and $g_{i}$ can be derived from the two
conditions (\ref{eq:orthoFG}) and (\ref{eq:connectionFG}).
Introducing equation~(\ref{eq:connectionFG}) into equation~(\ref{eq:orthoFG}) gives
\begin{equation}
  {\tilde \sigma}_{i}^{2} \, <g_{i},g_{j}> \, = \,
           <Lf_{i}, Lf_{j}> \, = \, {\tilde \sigma}_{i}^{2} \, \delta_{i,j}
	\label{eq:orthoLFLG}
\end{equation}
After integrating the vector product $<Lf_{i}, Lf_{j}>$ by parts
one gets [4]
\begin{equation}
   <Lf_{i}, Lf_{j}> \, = <L^{H}Lf_{i}, f_{j}> \, = \, <L^{F}Lf_{i}, f_{j}>
          + \, {\underline \beta}_{\rm L}[f_{i},f_{j}] \; ,
   \label{eq:integratedLfLf}
\end{equation}
where $L^{H}$ is the adjoint of $L$, $L^{F}$ the formal or Lagrange adjoint of $L$,
and ${\underline \beta}_{\rm L}[f_{i},f_{j}]$ a boundary functional.
If the boundary functional is zero equations~(\ref{eq:orthoFG}), (\ref{eq:orthoLFLG}),
and (\ref{eq:integratedLfLf}) lead to the linear differential equation
\begin{equation}
   L^{F}Lf = {\tilde \sigma}^{2} \, f
     \label{eq:basicDiffEq}
\end{equation}
Equations~(\ref{eq:integratedLfLf}) and (\ref{eq:basicDiffEq}) are the fundamental
equations for the derivation of the expressions for the analytical normal modes
in the limit of infinitely fine segmentation.
The boundary conditions which are required to derive specific infinite
sets $\{ f_{i} \}$ of analytical normal modes and the corresponding sets
$\{ {\tilde \sigma}_{i} \}$ of analytical singular values 
from the general solution of equation~(\ref{eq:basicDiffEq})
are obtained from the requirement that the boundary functional vanishes.
The form of the operator $L^{F}$ and the boundary functional
depend on the operator $L$. One of the
problems is therefore to find the correct expression for the linear differential
operator $L$ which corresponds, in the limit of infinitely fine segmentation,
to the matrix ${\bf A}$.
\section{One-dimensional segmentation}
\label{sec:1DimSolutions}
The method outlined in section~\ref{sec:introMethod} is first applied to the
mathematically easier case of one-dimensional segmentation. In addition, some of
the features appearing in this section are similar to the ones occuring in the
more interesting examples of two-dimensional segmentation discussed in
sections~\ref{sec:analytical2Dim} and \ref{sec:SVD2Dim}.
\subsection{SVD solutions}
  \label{sec:svd1Dim}
\begin{figure}[h]
     \centerline{\hbox{
	 \psfig{figure=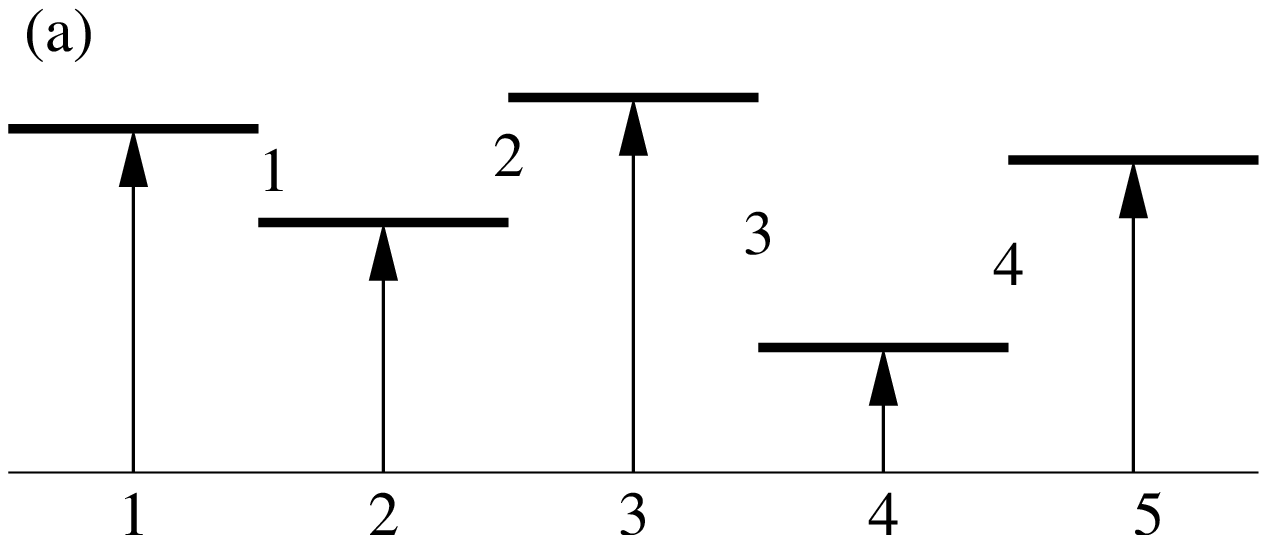,width=70mm}
         \psfig{figure=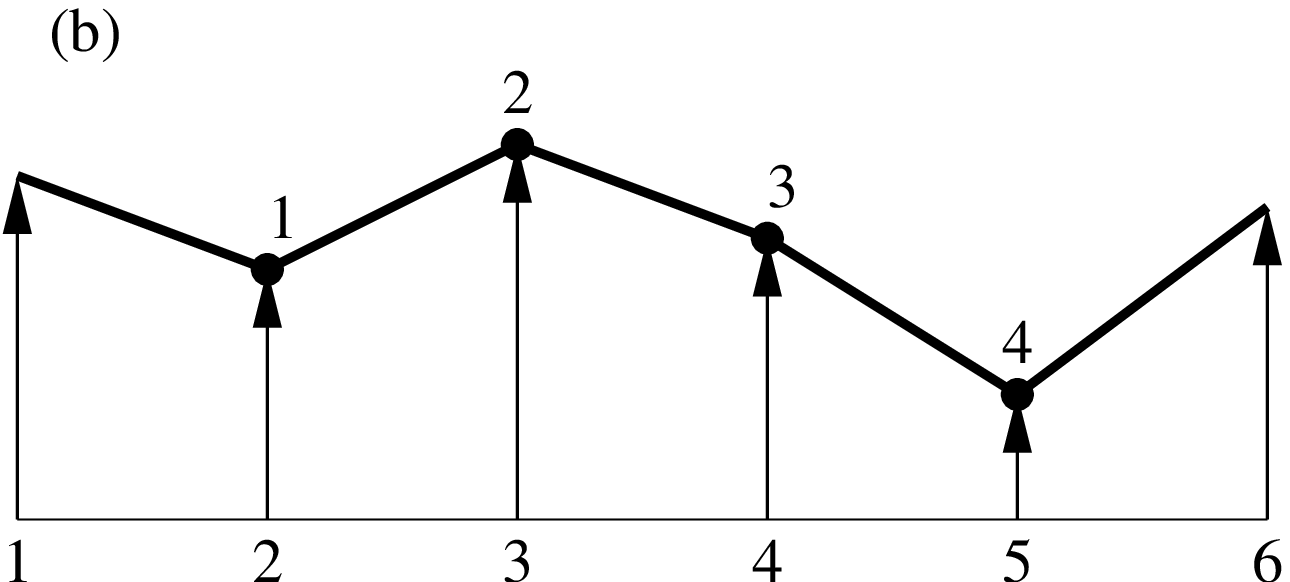,width=70mm}}}
   \caption{\label{fig:figure1} {\small {\bf (a) :} Segments with piston movements only,
          {\bf (b) :} Segments linked at the interfaces.}}
\end{figure}

Figure~\ref{fig:figure1} shows two examples which are based on linear chains of
segments.
In figure~\ref{fig:figure1}a each segment can only be moved in vertical
direction. If the actuator positions are denoted by $z_{i}$, the signals $s_{i}$
measured by vertical displacement sensors are given by
\begin{equation}
  s_{i} = z_{i+1} - z_{i}
      \label{eq:sensorSignalPiston}
\end{equation}
In figure~\ref{fig:figure1}b the segments are linked and the actuators are
at the links and at both ends of the chain. The signals $s_{i}$ are
defined by
\begin{equation}
  s_{i} = z_{i} - 2z_{i+1} + z_{i+2}
      \label{eq:sensorSignalBending}
\end{equation}
For small angles $s_{i}$ is approximately proportional to the difference between the
tilts of the segments. The matrices linking the actuator positions to the signals
are given by
\begin{equation}
  {\bf A_{{\rm a}}} = \left(
   \begin{array}{rrrrr}
   -1 & 1 & 0 & 0 & 0 \\
    0 & 1 &-1 & 0 & 0 \\
    0 & 0 & 1 &-1 & 0 \\
    0 & 0 & 0 & 1 &-1 \\
  \end{array}  \right )
\end{equation}
for the example (a) with $n_{{\rm sig}} = n_{{\rm act}} - 1$ and
\begin{equation}
  {\bf A_{{\rm b}}} = \left(
   \begin{array}{rrrrrr}
    1 &-2 & 1 & 0 & 0 & 0 \\
    0 & 1 &-2 & 1 & 0 & 0 \\
    0 & 0 & 1 &-2 & 1 & 0 \\
    0 & 0 & 0 & 1 &-2 & 1 \\
  \end{array}  \right )
\end{equation}
for the example (b) with $n_{{\rm sig}} = n_{{\rm act}} - 2$.
In both examples the number $n_{{\rm nz}}$ of normal modes with non-zero singular values
is equal to the number of signals.

\begin{figure}[h]
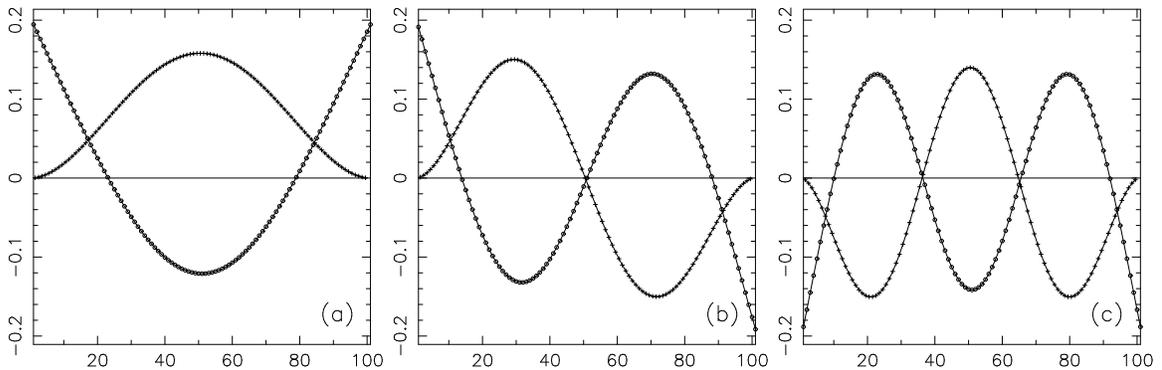

     \centerline{\hbox{
      \psfig{figure=figure2a.ps,width=50mm}
      \psfig{figure=figure2b.ps,width=50mm}
      \psfig{figure=figure2c.ps,width=50mm}}}
   \caption{\label{fig:figure2} {\small Example (b) : Actuator values
    (circles) and sensor signals (crosses) for the three SVD modes with the lowest
    singular values.
    The solid lines are the mode shapes calculated analytically for
    an infinite number of segments.}}
\end{figure}

For large values of $n_{{\rm act}}$ the lowest modes, which are the ones with the smallest
singular values, approach smooth functions.
Figure~\ref{fig:figure2} shows for the example (b) with 100 segments the three modes
with the lowest singular values.
\subsection{Analytical solutions for one-dimensional segmentation}
\subsubsection{Example (a)}
\label{sec:1DimPiston}
In the limit of infinitely small segments the operator $L$ which corresponds to
the matrix ${\bf A}_{{\rm a}}$ of example (a) is given by
\begin{equation}
  L = \frac{{\rm d}}{{\rm d}x} = \frac{1}{D} \, \frac{{\rm d}}{{\rm d}\xi}
                          \label{eq:diffOperExa}
\end{equation}
Then, with $-D$ and $D$ being the left and the right end of the chain of segments,
equation~(\ref{eq:integratedLfLf}) becomes
\begin{eqnarray}
  <Lf_{i}(x), Lf_{j}(x)> & = & \frac{1}{2D^{2}}
      \, \int_{-1}^{+1} \left( \frac{{\rm d}}{{\rm d}\xi}f_{i}(\xi) \right)
                    \left( \frac{{\rm d}}{{\rm d}\xi}f_{j}(\xi)\right) \, d\xi
            \label{eq:calcLHExa1} \\
   & = & - \frac{1}{2D^{2}}
      \, \int_{-1}^{+1}
        \left( \frac{{\rm d}^{2}}{{\rm d}\xi^{2}}f_{i}(\xi)\right) f_{j}(\xi) \, d\xi\,
        + \, \frac{1}{2D^{2}} \,
      \left[\left( \frac{{\rm d}}{{\rm d}\xi} f_{i}(\xi)\right) f_{j}(\xi) \right]_{-1}^{+1}
            \label{eq:calcLHExa2} 
\end{eqnarray}
For an arbitrary choice of a function
$f_{j}$ from the set $\{ f_{j} \} $ the boundary functional vanishes only if
\begin{equation}
    \left( \frac{{\rm d}}{{\rm d}\xi}f(\xi) \right)_{\xi = -1}
  = \left( \frac{{\rm d}}{{\rm d}\xi}f(\xi) \right)_{\xi = +1} = 0
  \label{eq:bcPiston1Dim}
\end{equation}
The differential equation~(\ref{eq:basicDiffEq}) in terms of the normalised variable
$\xi$ is then given by
\begin{equation}
  -\frac{1}{D^{2}} \, \frac{{\rm d}^{2}}{{\rm d}\xi^{2}} f(\xi) 
            = {\tilde \sigma}^{2} f(\xi),
                  \label{eq:diffEqExa}
\end{equation}
or, with the new normalised parameter $\sigma = D {\tilde \sigma}$ which has the advantage
that it is independent of the length $2D$ of the chain and has got the unit 1, by
\begin{equation}
  - \frac{{\rm d}^{2}}{{\rm d}\xi^{2}} f(\xi) 
            = \sigma^{2} f(\xi)
                  \label{eq:diffEqExaNorm}
\end{equation}
The general solutions are either symmetric or antisymmetric :
\begin{eqnarray}
   f_{{\rm s}}(\xi) & = & c_{{\rm s}} \cos(\sigma \xi)
           \label{eq:solSymmetricExa}\\
   f_{{\rm a}}(\xi) & = & c_{{\rm a}} \sin(\sigma \xi)
           \label{eq:solAsymmetricExa}
\end{eqnarray}
with $c_{{\rm s}}$ and $c_{{\rm a}}$ denoting arbitrary constants.
Introducing these equations into the boundary conditions (\ref{eq:bcPiston1Dim})
gives the equations from which the infinite sets of normalised analytical
singular values $\sigma_{i} = D {\tilde \sigma}_{i}$
for the symmetric and the antisymmetric cases can be obtained.
The $\sigma_{i}$ are identical to the wavenumbers of the solution functions
and given by
\begin{equation}
  \sigma_{i} = i\; \frac{\pi}{2}
    \label{eq:singValue1DimExa}
\end{equation}
Obviously, for the symmetric and antisymmetric cases the sets
$\{ f_{{\rm s}/{\rm a},i} \}$ and $\{ Lf_{{\rm s}/{\rm a},i} \}$,
representing the analytical position and signal modes respectively,
are orthogonal sets of functions.
The following relationships, which follow from the equations~(\ref{eq:orthoFG}),
(\ref{eq:orthoLFLG}), and (\ref{eq:calcLHExa1}) will be required in
section~\ref{sec:comp1DimExa} :
\begin{eqnarray}
  \sigma_{i}^{2} = D^{2} \, {\tilde \sigma}_{i}^{2}
   & = & D^{2} \, \frac{<Lf_{i}(x), Lf_{i}(x)>}{<f_{i}(x), f_{i}(x)>}
               \nonumber \\
   & = & \int_{-1}^{+1} \left( \frac{{\rm d}}{{\rm d}\xi}f_{i}(\xi) \right)^{2} \, d\xi
          \, {\big /} \, \int_{-1}^{+1} \left( f_{i}(\xi) \right)^{2} \, d\xi
    \label{eq:defSigmai}
\end{eqnarray}
\subsubsection{Example (b)}
\label{sec:1DimTilt}
In the limit of infinitely small segments the operator $L$ which correspnds to the
matrix ${\bf A}_{{\rm b}}$ of example (b) is given by
\begin{equation}
  L = \frac{{\rm d}^{2}}{{\rm d}x^{2}}
    = \frac{1}{D^{2}} \, \frac{{\rm d}^{2}}{{\rm d}\xi^{2}}
                          \label{eq:diffOperExb}
\end{equation}
Taking the same steps as in section~\ref{sec:1DimPiston} one gets
\begin{eqnarray}
  L^{{\rm F}}L & = & \frac{1}{D^{4}}\frac{{\rm d}^{4}}{{\rm d}\xi^{4}}
                    \label{eq:LFExb}\\
  \underline{\beta}_{L}[f_{i},f_{j}] & = & \frac{1}{2D^{4}}
         \left[ \left( \frac{{\rm d}^{2}}{{\rm d}\xi^{2}} f_{i}\right)
                \frac{{\rm d}f_{j}}{{\rm d}\xi}
        - \left( \frac{{\rm d}^{3}}{{\rm d}\xi^{3}} f_{i} \right)
                 f_{j} \right]_{-1}^{+1}
               \label{eq:boundaryFunctionalExb}
\end{eqnarray}
The differential equation is therefore given by
\begin{equation}
  (\frac{{\rm d}^{4}}{{\rm d}\xi^{4}} - \sigma^{2})\; f(\xi)  = 0
                  \label{eq:diffEqExb}
\end{equation}
with $\sigma = D^{2}{\tilde \sigma}$.
The fourth-order differential operator in equation~(\ref{eq:diffEqExb})
can be factorised into two second-order differential operators :
\begin{equation}
  (\frac{{\rm d}^{2}}{{\rm d}\xi^{2}} - \sigma)
  (\frac{{\rm d}^{2}}{{\rm d}\xi^{2}} + \sigma)\; f(\xi) = 0
                  \label{eq:diffEqExb1}
\end{equation}
With the definition
\begin{equation}
  \lambda^{2} = \sigma
    \label{eq:defLambda}
\end{equation}
the symmetric and antisymmetric
solutions are, with $a_{{\rm s}}$ and $a_{{\rm a}}$ denoting additional
coefficients,
\begin{eqnarray}
   f_{{\rm s}}(\xi) & = & c_{{\rm s}} [\cos(\lambda \xi)
                       + a_{{\rm s}}\cosh(\lambda \xi)]
           \label{eq:solSymmetricExb}\\
   f_{{\rm a}}(\xi) & = & c_{{\rm a}} [\sin(\lambda \xi)
                        + a_{{\rm a}}\sinh(\lambda \xi)]
           \label{eq:solAsymmetricExb}
\end{eqnarray}
Both sets of solutions require two boundary conditions which can be derived from
the boundary functional~ (\ref{eq:boundaryFunctionalExb}). This must be zero
for arbitrary functions taken from the sets of functions $\{ f_{{\rm s},j} \}$
or $\{ f_{{\rm a},j} \}$. Since at least for some of them both the function and
its derivative are non-zero at the edges the four conditions which guarantee that
the boundary functional is zero for all combinations of $i$ and $j$ are
\begin{equation}
   \left( \frac{{\rm d}^{2}}{{\rm d}\xi^{2}}f(\xi) \right)_{\xi = -1}
 = \left( \frac{{\rm d}^{2}}{{\rm d}\xi^{2}}f(\xi) \right)_{\xi = +1} = 0
    \label{eq:BC1Exb}
\end{equation}
\begin{equation}
   \left( \frac{{\rm d}^{3}}{{\rm d}\xi^{3}}f(\xi) \right)_{\xi = -1}
 = \left( \frac{{\rm d}^{3}}{{\rm d}\xi^{3}}f(\xi) \right)_{\xi = +1} = 0
       \label{eq:BC2Exb}
\end{equation}
Introducing the general solutions (\ref{eq:solSymmetricExb}) and 
(\ref{eq:solAsymmetricExb}) into equations~(\ref{eq:BC1Exb}) and (\ref{eq:BC2Exb})
gives the equations from which the wavenumbers $\lambda_{i}$ and the coefficients
$a_{{\rm s},i}$ and $a_{{\rm a},i}$ can be calculated.
The wavenumber $\lambda_{i}$ of a mode $i$
is approximately given by
\begin{equation}
  \lambda_{i} = \frac{\pi}{4}\, (1 + 2i)
    \label{eq:singValue1DimExb}
\end{equation}
This approximation is very accurate for large $i$. For small $i$ the differences are
only of the order of a few percent.
According to equation (\ref{eq:defLambda}) the normalised analytical singular values
$\sigma_{i}$ are the squares of the wavenumbers $\lambda_{i}$.
\subsection{Comparison between SVD-modes and analytical modes}
\subsubsection{Example (a)}
\label{sec:comp1DimExa}
\begin{figure}[h]
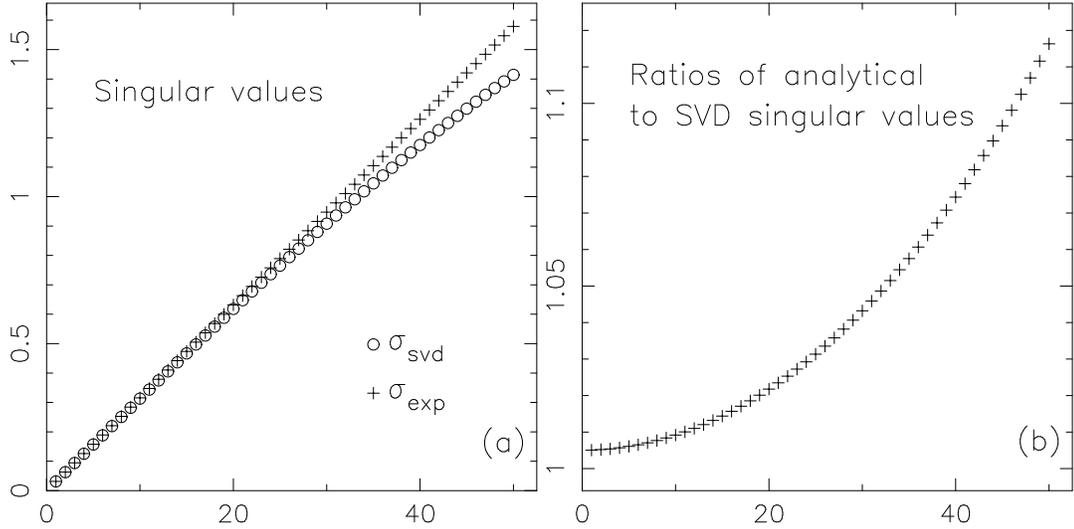

     \centerline{\hbox{
      \psfig{figure=figure3a.ps,width=70mm}
      \psfig{figure=figure3b.ps,width=70mm}}}
   \caption{\label{fig:figure3} {\small Example (a).
          {\bf (a) :} Singular values calculated analytically
          (crosses) and by SVD (circles), {\bf (b) :} Ratios of these
          values for the lowest 50 modes.}}
\end{figure}

One can derive an expression for the singular values $\sigma_{{\rm svd,an},i}$
expected from SVD based on the normalised analytical singular values $\sigma_{i}$.
From the equations~(\ref{eq:orthoSetsUV}) and (\ref{eq:connectionUV}) one gets
\begin{eqnarray}
 \sigma_{{\rm svd},i}^{2} & = &
   \frac{<{\bf A}{\bf v}_{i},{\bf A}{\bf v}_{i}>}{<{\bf u}_{i},{\bf u}_{i}>}
     = \frac{<{\bf A}{\bf v}_{i},{\bf A}{\bf v}_{i}>}{<{\bf v}_{i},{\bf v}_{i}>}
         \nonumber \\
   & = & \frac{n_{{\rm sig}}}{n_{{\rm act}}} \,
       \frac{<{\bf A}{\bf v}_{i},{\bf A}{\bf v}_{i}>}{n_{{\rm sig}}} \, {\big /} \,
        \frac{<{\bf v}_{i},{\bf v}_{i}>}{n_{{\rm act}}} \nonumber \\
   & = & \frac{n_{{\rm sig}}}{n_{{\rm act}}} \,
          \frac{\epsilon_{{\rm sig},i}^{2}}{\epsilon_{{\rm act},i}^{2}} \, ,
            \label{eq:sigmaSvdiSq}
\end{eqnarray}
where $\epsilon_{{\rm sig},i}$ and $\epsilon_{{\rm act},i}$ are the r.m.s. values of the
signals and the actuator positions, respectively.
To derive an expression for $\epsilon_{{\rm sig},i}$ let $d$ be the length of a segment,
$n_{{\rm seg}}$ the number of segments,
and $2D = n_{{\rm seg}}d$ the length of the
full linear chain. If $f_{i}(x)$ is one of the analytical normal modes going through the
centers of the segments and the $x_{j}$ are the locations of the sensors,
the mean square $\epsilon_{{\rm sig},i}^{2}$ of the signals is approximately given by
\begin{equation}
  \epsilon_{{\rm sig},i}^{2} = d^{2} \, \frac{1}{n_{{\rm sig}}} \sum_{j=1}^{n_{{\rm sig}}}
          \left(\frac{{\rm d}f_{i}(x)}{{\rm d}x}\right)^{2}_{x=x_{j}}
        \label{eq:meanSquareExA1}
\end{equation}
For a very large number of segments the averaged sum can be replaced by the integral
expressions in equation~(\ref{eq:calcLHExa1}) with $i=j$, leading to the analytical mean
square $\epsilon_{{\rm sig,an},i}^{2}$ of the signals :
\begin{equation}
  \epsilon_{{\rm sig,an},i}^{2}  =
        \frac{1}{2} \left(\frac{d}{D} \right)^{2} \, \int_{-1}^{+1}
           \left(\frac{{\rm d}f_{i}(\xi)}{{\rm d}\xi}\right)^{2} {\rm d}\xi
          \label{eq:meanSquareSensorsExA}
\end{equation}
Dividing equation~(\ref{eq:meanSquareSensorsExA}) by the analytical mean square
$\epsilon_{{\rm act,an},i}^{2}$ of the positions
\begin{equation}
   \epsilon_{{\rm act,an},i}^{2} = \frac{1}{2} \, \int_{-1}^{+1} f_{i}^{2}(\xi) \, {\rm d}\xi
        \label{eq:defEpsilonAct}
\end{equation}
and using equation~(\ref{eq:defSigmai}) gives
\begin{eqnarray}
 \frac{\epsilon_{{\rm sig,an},i}^{2}}{\epsilon_{{\rm act,an},i}^{2}}
& = & \left(\frac{d}{D} \right)^{2} \, \int_{-1}^{+1}
           \left(\frac{{\rm d}f_{i}(\xi)}{{\rm d}\xi}\right)^{2} {\rm d}\xi \;\;
      {\Big  /} \;\;
            \int_{-1}^{+1} f_{i}^{2}(\xi) \, {\rm d}\xi   \nonumber \\
    & = & \left(\frac{d}{D} \right)^{2} \, \sigma_{i}^{2}
       = d^{2} \, {\tilde \sigma}_{i}^{2}
           \label{eq:rmsSensorsDivRmsFunctionExa}
\end{eqnarray}
Introducing equation~(\ref{eq:rmsSensorsDivRmsFunctionExa}) into
equation~(\ref{eq:sigmaSvdiSq}) and taking the square root gives an expression
not any more for the true singular values $\sigma_{{\rm svd},i}$ but for the ones
$\sigma_{{\rm svd,an},i}$ based on the analytical approximation with an infinite
number of segments.
The relationship between the normalised analytical singular
value $\sigma_{i}$ of the mode $i$ and the corresponding one $\sigma_{{\rm svd,an},i}$
expected from SVD is then
\begin{equation}
  \sigma_{{\rm svd,an},i} = \sqrt{\frac{n_{{\rm sig}}}{n_{{\rm act}}}} \,
                        \frac{d}{D} \, \sigma_{i}
      \label{eq:svdAnalytical}
\end{equation}
For a chain with 100 segments figure~\ref{fig:figure3}a shows the singular values
$\sigma_{{\rm svd,an},i}$ according to equation~(\ref{eq:svdAnalytical}) (crosses)
and the true singular values
$\sigma_{{\rm svd},i}$ calculated by SVD (circles).
As one would expect, they only coincide for the lowest smooth modes. For the higher
modes the analytical calculation overestimates the singular values.
This can also be seen
in figure~\ref{fig:figure3}b which shows the ratios of
$\sigma_{{\rm svd,an},i}$ to $\sigma_{{\rm svd},i}$.
Also for the lowest modes these ratios are not exactly equal
to 1, but approximately equal to $1 + 0.5/n_{{\rm seg}}$.
For a large number of segments the shapes of the SVD solutions with the smallest singular
values are in perfect agreement with the ones of the analytical solutions.
Since the normal modes are used for the alignment of segmented mirrors, a common
question is how the signal errors propagate to actuator errors, that is, which r.m.s.
$\epsilon_{{\rm act,n}}$ of the actuator errors is generated by noise in the signals
with a given r.m.s. of $\epsilon_{{\rm sig,n}}$.
The r.m.s. $\epsilon_{{\rm act,n}}$ can be calculated by
a sum involving the $n_{{\rm nz}}$ non-zero singular values.
Two r.m.s. values will be calculated. An analytical one denoted by $\epsilon_{{\rm act,n,an}}$,
where the sum is formed with the singular values $\sigma_{{\rm svd,an},i}$, and a true one
denoted by $\epsilon_{{\rm act,n,svd}}$, where the sum is formed with the true singular
values $\sigma_{{\rm svd},i}$. In general, the mean square
$\epsilon_{{\rm act,an}}^{2}$ of the actuator errors based on analytical singular values
is the sum of the mean squares $\epsilon_{{\rm act,an},i}^{2}$ of the individual modes.
Using also equation~(\ref{eq:rmsSensorsDivRmsFunctionExa}) one gets
\begin{equation}
    \epsilon_{{\rm act,an}}^{2} = \sum_{i=1}^{n_{{\rm nz}}} \epsilon_{{\rm act,an},i}^{2}
         = \sum_{i=1}^{n_{{\rm nz}}} \epsilon^{2}_{{\rm sig,an},i}
             \, \left(\frac{D}{d} \right)^{2}
                 \frac{1}{\sigma_{i}^{2}}
           \label{eq:epsilonActAn}
\end{equation}
If the signal errors are random, 
each of the $n_{{\rm sig}}$ signal modes contributes an equal
amount $\epsilon_{{\rm sig,n},i}=\epsilon_{{\rm sig,n}}/\sqrt{n_{{\rm sig}}}$
to the total r.m.s. $\epsilon_{{\rm sig,n}}$ of the signal noise.
The analytically expected total mean square $\epsilon_{{\rm act,n,an}}^{2}$ of the actuator
errors due to random signal noise is then, after replacing
$\epsilon_{{\rm sig,an},i}$ by $\epsilon_{{\rm sig,n},i} =
\epsilon_{{\rm sig,n}}/\sqrt{n_{{\rm sig}}}$ and $\sigma_{i}$ by the explicit expression
in equation~(\ref{eq:singValue1DimExa}), given by
\begin{equation}
  \epsilon_{{\rm act,n,an}}^{2}  = 
         \epsilon^{2}_{{\rm sig,n}} \,
             \frac{1}{\pi^{2}} \, \left(\frac{2D}{d} \right)^{2} \, 
         \frac{1}{n_{{\rm sig}}} \,
          \sum_{i=1}^{n_{{\rm nz}}} \frac{1}{i^{2}}
           \label{eq:totalRmsActNoiseExa1}
\end{equation}
For a large number of segments $n_{{\rm sig}}$ can be replaced by $2D/d$.
Furthermore, because of its fast convergence, the sum
from 1 to $n_{{\rm nz}}$ is, for large $n_{{\rm nz}}$, approximately equal to the sum
from 1 to infinity, with a value of $\pi^{2}/6$. All this finally gives
\begin{equation}
  \epsilon_{{\rm act,n,an}} \approx
       \epsilon_{{\rm sig,n}} \; \sqrt{\frac{1}{6}} \, \sqrt{\frac{2D}{d}} 
           \label{eq:totalRmsActNoiseExa}
\end{equation}
For a given r.m.s. of the signal noise the r.m.s. of the actuator error therefore
scales, as one would expect, with the square root of the number of segments.
Equation~(\ref{eq:totalRmsActNoiseExa1}) can also be written as
\begin{equation}
  \epsilon_{{\rm act,n,an}}^{2} = \epsilon_{{\rm sig,n}}^{2} \, \frac{1}{n_{{\rm act}}}
        \sum_{i=1}^{n_{{\rm nz}}} \frac{1}{\sigma_{{\rm svd,an},i}^{2}}
	\label{eq:totalRmsActNoiseExa1Mod}
\end{equation}
Similarly, the true r.m.s. $\epsilon_{{\rm act,n,svd}}$ calculated with the true SVD singular
values $\sigma_{{\rm svd},i}$ is given by
\begin{equation}
  \epsilon_{{\rm act,n,svd}}^{2} = \epsilon_{{\rm sig,n}}^{2} \, \frac{1}{n_{{\rm act}}}
        \sum_{i=1}^{n_{{\rm nz}}} \frac{1}{\sigma_{{\rm svd},i}^{2}}
	\label{eq:totalRmsActNoiseExa1Svd}
\end{equation}
The result from equation~(\ref{eq:totalRmsActNoiseExa1Svd}) is effectively identical
to the one from equation~(\ref{eq:totalRmsActNoiseExa}).
The same result is also obtained by simple statistical
considerations, that is by summing up random piston movements along a line of segments
and calculating the r.m.s. after subtracting the average.
\subsubsection{Example (b)}
\begin{figure}[h]
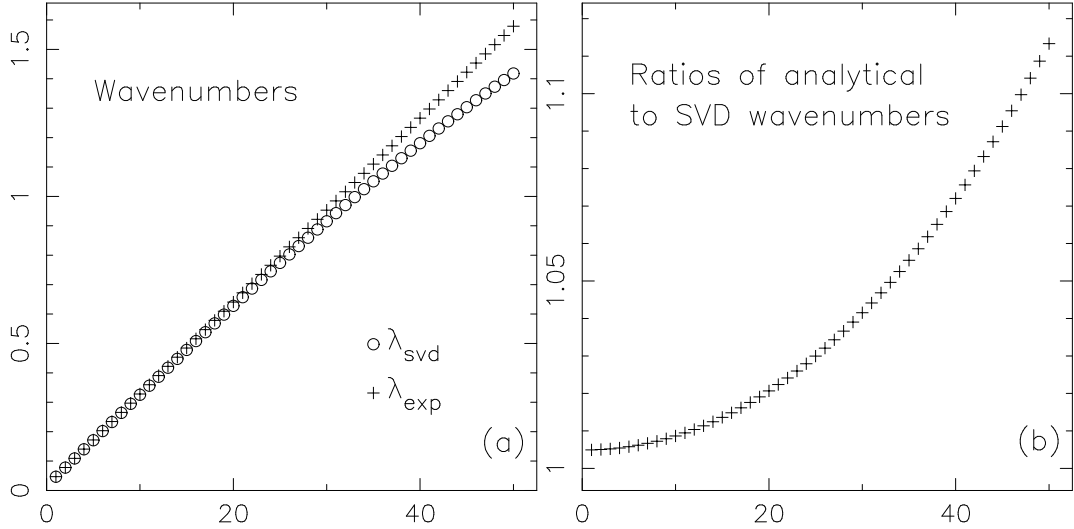

     \centerline{\hbox{
      \psfig{figure=figure4a.ps,width=70mm}
      \psfig{figure=figure4b.ps,width=70mm}}}
   \caption{\label{fig:figure4} {\small Example (b).
          {\bf (a) :}
          Wavenumbers calculated analytically
          (crosses) and by SVD (circles), {\bf (b) :}
          Ratios of these values for the lowest 50 modes.}}
\end{figure}

The expression for $\sigma_{{\rm svd,an},i}$ can be derived in the same way as
in section~\ref{sec:comp1DimExa}. One gets for a specific mode $i$
\begin{eqnarray}
  \left( \frac{\epsilon_{{\rm sig,an},i}}{\epsilon_{{\rm act,an},i}} \right)^{2} & = &
          \frac{d^{2}}{D^{4}}  \, \int_{-1}^{+1}
           \left(\frac{{\rm d}f_{i}(\xi)}{{\rm d}\xi}\right)^{4}\, {\rm d}\xi \;\;
           {\Big /} \;\;
            \int_{-1}^{+1} f_{i}^{2}(\xi) {\rm d}\xi \\
    & = & \frac{d^{2}}{D^{4}} \, \sigma_{i}^{2}
      =  d^{2} \, {\tilde \sigma}_{i}^{2} 
           \label{eq:rmsSensorsDivRmsFunctionExb}
\end{eqnarray}
with $\sigma_{i} = D^{2}{\tilde \sigma}_{i}$.
Here $\epsilon_{{\rm sig,an},i}$ is the r.m.s. of the angular signal error in radians.
The singular value $\sigma_{{\rm svd,an},i}$ expected from SVD is then related to
the corresponding analytical singular value $\sigma_{i}$ by
\begin{equation}
  \sigma_{{\rm svd,an},i} = \sqrt{\frac{n_{{\rm sig}}}{n_{{\rm act}}}} \,
                        \frac{d}{D^{2}} \, \sigma_{i}
\end{equation}
For a chain with 100 segments figure~\ref{fig:figure4}a shows the square roots
of the singular values $\sigma_{{\rm svd,an},i}$ (crosses), that is the wavenumbers
corresponding to $\sigma_{{\rm svd,an},i}$ according to equation~(\ref{eq:defLambda}),
and of the true ones $\sigma_{{\rm svd},i}$ (circles).
Figure~\ref{fig:figure4}a for the wavenumbers is very similar to
the corresponding figure~\ref{fig:figure3}a for the singular values
of example (a).
Also the figure \ref{fig:figure4}b for the ratios of the square roots
of $\sigma_{{\rm svd,an},i}$ to the square roots of $\sigma_{{\rm svd},i}$ is effectively
identical to the corresponding figure \ref{fig:figure3}b of example (a)
for the ratios of the singular values.
The total mean square of the actuator errors is given by, using also
equation~(\ref{eq:singValue1DimExb}) for the analytical singular values,
\begin{equation}
  \epsilon_{{\rm act,n,an}}^{2} = \epsilon^{2}_{{\rm sig,n}} \,
           \left(\frac{4}{\pi}\right)^{4} \,
              \frac{D^{4}}{d^{2}} \,
         \frac{1}{n_{{\rm sig}}} \,
          \sum_{i=1}^{n_{{\rm nz}}} \left(\frac{1}{1+2i}\right)^{4}
           \label{eq:totalRmsActNoiseExb1}
\end{equation}
Next, $n_{{\rm sig}}$ is replaced by $2D/d$. Furthermore, because of its fast convergence,
the sum from 1 to $n_{{\rm nz}}$ is, for large $n_{{\rm nz}}$, approximately equal
to the sum from 1 to infinity, with a value of $\pi^{4}/120 - 1$. All this finally gives
\begin{equation}
  \epsilon_{{\rm act,n,an}} \approx
        0.0982 \, D \, \sqrt{\frac{2D}{d}} \, \epsilon_{{\rm sig,n}}
           \label{eq:totalRmsActNoiseExb}
\end{equation}
For a chain of 100 segments the difference
between the true value $\epsilon_{{\rm act,n,svd}}$ obtained from
equation~(\ref{eq:totalRmsActNoiseExa1Svd}) applied to example (b)
and $\epsilon_{{\rm act,n,an}}$ from
equation~(\ref{eq:totalRmsActNoiseExb}) is less than 1\%.
The solid lines in figure~\ref{fig:figure2} show the analytical solutions with the
three smallest singular values. Clearly, they are in perfect agreement with the SVD
solutions.
\section{Analytical solutions for two-dimensional segmentation}
      \label{sec:analytical2Dim}
\subsection{General remarks}
\begin{figure}[h]
     \centerline{\hbox{
      \psfig{figure=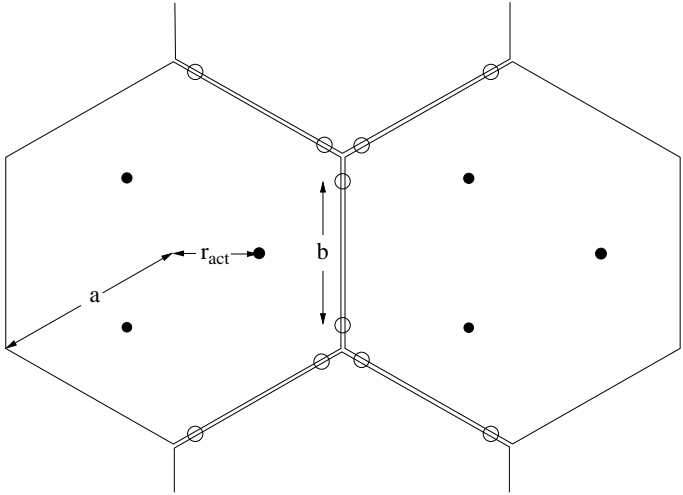,width=90mm}}}
   \caption{\label{fig:figure5} {\small Locations of the sensors at the
    intersegment edges of hexagonal segments. The filled circles in the segments
    mark the positions of the actuators.}}
\end{figure}

There are several combinations of signals and actuator positions which could be
studied. In this chapter we discuss mainly two examples which are of particular interest
for the alignment and phasing of segmented mirrors. We assume that
the segments are hexagonal and the signals are measured at locations along the
intersegment edges as shown in figure~\ref{fig:figure5}.
In both examples the signals are the relative
vertical displacements between adjacent segments at the locations of the sensors.
Such signals can be obtained by optical measurements or by position sensors.
In principle, the signals can also include the relative tilts between adjacent segments,
defined here as a change in the slope of the segments perpendicular to the edges
at the locations of the sensors.
Usually, the signals generated by relative tilts are much smaller than signals generated
by relative vertical displacements [6]. Except for one particular mode, the defocus mode,
which is discussed in section~\ref{sec:relativeTilt},
relative tilts at the sensors are therefore of less significance.

The two examples differ in the number of the degrees of freedom
for the movements of the segments.
\begin{itemize}
\vspace{-3mm}
\item {\it Piston movements}\\
In this case, which is the two-dimensional equivalent of
the one-dimensional example (a) presented in section~\ref{sec:1DimPiston},
the segments can only move in piston and the positions of the three
actuators of one segment are therefore all identical.
The signals along one intersegment edge are
all the same and it would be sufficient to consider just one signal
along one edge.
This example describes a situation where the segments have already
been perfectly aligned in tilt and the remaining phasing errors are corrected
by pure piston movements of the segments, a procedure which is used for the phasing
of the segmented primary mirror of the Keck telescope based on optical measurements.
\item {\it Piston, tip and tilt movements}\\
In this case, which is to some extent related to the one-dimensional example (b)
presented in section~\ref{sec:1DimTilt}, there are three degrees of freedom for
the movements of the segments, namely piston, tip and tilt.
It describes situations where a full correction of the misalignments including phasing
is based on relative vertical displacement signals at the locations shown in
figure~\ref{fig:figure5}. It is used for the fast stabilization of the
alignment in the Keck telescope with signals obtained by position sensors.
\end{itemize}
The sensors at the edges of the hexagonal segments detect signals in the three
directions  $\theta_{1}=-60^{\circ}$,
$\theta_{2}=0^{\circ}$ and $\theta_{3}=60^{\circ}$ with respect to the horizontal
axis in figure~\ref{fig:figure5}.
The derivative of a function $f(r,\varphi)$ in one of the directions $\theta_{k}$,
$k=1,2,3$, 
expressed as derivatives with respect to the radial and azimuthal coordinates $r$
and $\varphi$ of a cylindrical coordinate system, is given by
\begin{equation}
T_{k}\, f(r,\varphi) = 
            \frac{\partial}{\partial s_{\theta_{k}}}\, f(r,\varphi)
                     =  \frac{1}{R} \,
           \left( \cos(\varphi - \theta_{k})\frac{\partial}{\partial \rho} 
       - \frac{\sin(\varphi - \theta_{k})}{\rho}\frac{\partial}{\partial \varphi}
                \right) \, f(\rho,\varphi) \; ,
          \label{eq:firstDerivative}
\end{equation}
In the limit of infinitesimally small segments the lowest SVD modes always follow
rotational symmetries $n$. They can therefore be written as a product of a radial and an
azimuthal component :
\begin{equation}
  {f(r,\varphi)} = f_{n}(r) \, \cos(n\varphi)
  \label{eq:fTofn}
\end{equation}
The differential operators for curvatures are second derivatives based on
the operator in equation~(\ref{eq:firstDerivative}). Here we use the following approximation.
Strictly speaking, the curvature $K$ along a direction $s$ is defined by
\begin{equation}
 K = \frac{{\rm d}^{2}f}{{\rm d}s^{2}} \, {\Big /} \,
    \left[1+ \left(\frac{{\rm d}f}{{\rm d}s}\right)^{2} \right] ^{3/2}
\end{equation}
Since the derivative in the denominator is small compared to unity, we will replace
the denominator by 1. This is equivalent to approximating the spherical surface
by a paraboloid with a radial dependence on $\rho^{2}$.
\subsection{Piston movements}
\label{sec:2DimAnalyticalP}
The two-dimensional case which is equivalent to the one-dimensional example in
section~\ref{sec:1DimPiston}
is a configuration where the segments can only perform piston movements
and the sensors measure only the relative vertical displacements. In this case the
differential operator is given by the operator $T_{k}$ in equation~(\ref{eq:firstDerivative}).
The starting point for the derivation of the analytical normal modes is a vector product
averaged over the three directions $\theta_{k}$ :
\begin{equation}
      \frac{1}{3} \,
       \sum_{k=1}^{3} <T_{k}f_{i}(r,\varphi), T_{k}f_{j}(r,\varphi)>
      = \frac{1}{3} \; \frac{1}{\pi(1 - \rho_{1}^{2})}
              \,\sum_{k=1}^{3}\int {\rm d}\varphi
            \int_{\rho_{1}}^{1} \rho \, {\rm d}\rho
                \, (T_{k}f_{i}(\rho,\varphi)) \,
                       (T_{k}f_{j}(\rho,\varphi))
             \label{eq:vecProdPiston}
\end{equation}
After introducing equation~(\ref{eq:fTofn}),
integrating over the azimuth angle $\varphi$, and summing over $k$
equation~(\ref{eq:vecProdPiston}) becomes
\begin{equation}
    \frac{1}{3} \, \sum_{k=1}^{3} <T_{k}f_{i}(r,\varphi), T_{k}f_{j}(r,\varphi)> \,
               = \frac{1}{2R^{2}} \, \frac{1 + \delta_{n,0}}{1 - \rho_{1}^{2}} \;
                \int_{\rho_{1}}^{1}{ \rho \, {\rm d}\rho \, \left [ 
                \left( \frac{{\rm d}}{{\rm d}\rho}f_{n,i}(\rho) \right)
                \left( \frac{{\rm d}}{{\rm d}\rho}f_{n,j}(\rho) \right)
                 + \frac{n^{2}}{\rho^{2}}f_{n,i}(\rho)f_{n,j}(\rho) \right ]}
 \label{eq:vecProdPiston1}
\end{equation}
Following the procedure outlined in section~\ref{sec:introMethod} one gets
\begin{eqnarray}
  L^{F}L & = & -  \frac{1}{2R^{2}} \, L_{{\rm B},n}
         \label{eq:LFLPiston} \\
      \underline{\beta}_{{\rm L}}[f_{n,i},f_{n,j}] & = &
    \frac{1}{2R^{2}} \, \frac{1 + \delta_{n,0}}{1 - \rho_{1}^{2}}
     \left[ \left( \frac{{\rm d}}{{\rm d}\rho}f_{n,i}(\rho) \right)
          f_{n,j}(\rho)\right]_{\rho_{1}}^{1}
    \label{eq:boundaryFunctionalP}
\end{eqnarray}
where $L_{{\rm B},n}$ is the operator appearing in the Bessel differential equation :
\begin{equation}
  L_{{\rm B},n} = \frac{{\rm d}^{2}}{{\rm d}\rho^{2}} +
                   \frac{1}{\rho}\frac{{\rm d}}{{\rm d}\rho} -
                   \frac{n^{2}}{\rho^{2}}
\end{equation}
The general solution of
\begin{equation}
  L_{{\rm B},n}f_{n}(\rho) = - \sigma^{2}f_{n}(\rho)
         \label{eq:diffEqPiston}
\end{equation}
with $\sigma$ related to ${\tilde \sigma}$ in
equation~(\ref{eq:basicDiffEq}) by $\sigma = \sqrt{2}R{\tilde \sigma}$
is a sum of the Bessel functions $J_{n}(\sigma \rho)$ and $Y_{n}(\sigma \rho)$
with coefficients $c_{{\rm J}}$ and $c_{{\rm Y}}$ :
\begin{equation}
  f_{n}(\rho) = c_{{\rm J}}J_{n}(\sigma \rho) + c_{{\rm Y}}Y_{n}(\sigma \rho)
         \label{eq:genSolPiston}
\end{equation}
For arbitrary functions $f_{n,j}$ taken from the set of general solutions
(\ref{eq:genSolPiston}) the boundary functional (\ref{eq:boundaryFunctionalP}) vanishes
if the expression in the square brackets
is zero for both $\rho = \rho_{1}$ and $\rho = 1$, that is if the derivatives vanish
at both the inner and outer edges of the mirror.
Introducing the general solution (\ref{eq:genSolPiston}) into the two boundary conditions
gives a set of two linear homogeneous equation
in $c_{{\rm J}}$ and $c_{{\rm Y}}$ from which the 
infinite set of normalised analytical singular values $\sigma_{n,i}$
and the corresponding ratios $(c_{{\rm Y}}/c_{{\rm J}})_{n,i}$ can be obtained.
As in the corresponding one-dimensional example discussed in section~\ref{sec:1DimPiston}
the singular values $\sigma_{n,i}$ are identical to the wavenumbers of the functions,
in this case two Bessel functions,
contributing to the solution (\ref{eq:genSolPiston}).

\begin{figure}[h]
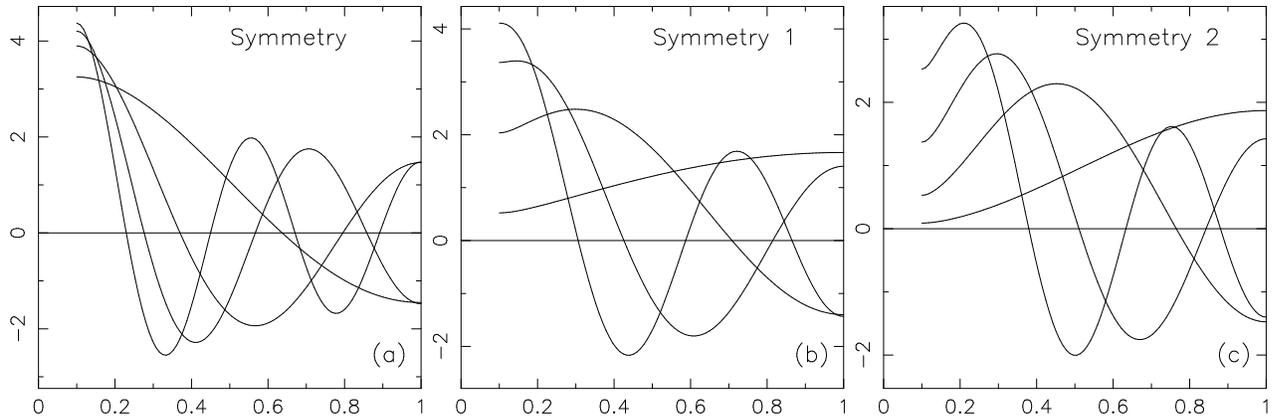

     \centerline{\hbox{
      \psfig{figure=figure6a.ps,width=55mm}
      \psfig{figure=figure6b.ps,width=55mm}
      \psfig{figure=figure6c.ps,width=55mm}}}
   \caption{\label{fig:figure6} {\small Lowest analytical
    normal modes of an annular mirror with $\rho_{1} = 0.1$ for pure piston movements
    for the rotational symmetries 0, 1, and 2.}}
\end{figure}

For the rotational symmetries 0, 1, and 2 figure~\ref{fig:figure6}
shows, for $\rho_{1} = 0.1$, the four analytical normal modes with the smallest singular values
$\sigma_{n,i}$. Like the Zernike polynomials the modes can be classified 
according to their rotational symmetries and,
within each rotational symmetry, according to their number of nodes, that is zeros
along the radial coordinate. The only mode, which is, compared with the set of Zernike
modes, missing in the set $\{f_{n,i}\}$
is the lowest mode of rotational symmetry 0 with no nodes. This reflects the fact
that a uniform piston movement of all segments does not generate relative vertical
displacements at intersegment edges.

The major difference to the Zernike polynomials is that, due to the boundary conditions
following from equation~(\ref{eq:boundaryFunctionalP}), the first derivatives are,
as in the one-dimensional example in section~\ref{sec:1DimPiston}, zero at the edges.
In the case of a completely filled mirror, that is $\rho{_1} = 0$, the function
$Y_{n}(\sigma \rho)$, which diverges at $\rho = 0$, drops out of the general solution,
which is then given by
\begin{equation}
  f_{n}(\rho) = c_{{\rm J}}J_{n}(\sigma \rho)
         \label{eq:genSolPistonNoHole}
\end{equation}
In this case the expression inside the square brackets of
the boundary functional~(\ref{eq:boundaryFunctionalP})
is automatically zero for $\rho = 0$. This leaves therefore
only one boundary condition which is sufficient to determine the infinite set of
normalised analytical singular values $\sigma_{n,i}$
which are also identical to the wavenumbers.
The functions look similar to the ones in
figure~\ref{fig:figure6}, but now, for all symmetries $n>0$, they
are zero at the centre of the mirror.
\subsection{Piston, tip and tilt movements}
\label{sec:2DimAnalyticalPTT}
\subsubsection{Differential operator}
As in section~\ref{sec:2DimAnalyticalP} the sensors detect only relative vertical
displacements, but the segments can now be controlled in piston, tip and tilt.
Let a smooth function over hexagonal areas be approximated by planes. For a spherical
surface this can be done without generating relative axial displacements.
The sensors therefore measure locally only the difference to a surface with, also locally,
a constant curvature in all directions. This is the same as the torsion of the surface
which is given by, using also equation~(\ref{eq:fTofn}),
\begin{eqnarray}
   L_{k}f(r,\varphi) & = &
  \frac{\partial^{2} f(r,\varphi)}{\partial s_{\theta_{k}} \partial s_{\theta_{k}-90^{\circ}}}
     \nonumber \\
      & = & \frac{1}{2R^{2}}\; \left[
            \sin(2(\varphi-\theta_{k})) \; \cos(n\varphi) \;\; L_{n,1}f_{n}(\rho)
            + 2\cos(2(\varphi-\theta_{k})) \; \sin(n\varphi)
                       \;\; L_{n,2}f_{n}(\rho) \right]
    \label{eq:operRPhiTheta}
\end{eqnarray}
with the two operators $L_{n,1}$ and $L_{n,2}$ defined by
\begin{eqnarray}
   L_{n,1} & = & \frac{{\rm d}^{2}}{{\rm d} \rho^{2}}
           - \frac{1}{\rho}\frac{{\rm d} }{{\rm d} \rho}
           + \frac{n^{2}}{\rho^{2}}
              \label{eq:lr1}\\
   L_{n,2} & = & - \frac{n}{\rho}\frac{{\rm d} }{{\rm d} \rho}
                 + \frac{n}{\rho^{2}}
              \label{eq:lr2}
\end{eqnarray}
\subsubsection{Derivation of the analytical normal modes}
\label{sec:derivModesPTT}
The procedure outlined in section~\ref{sec:introMethod} starts off with the following
vector product averaged over the three direction $\Theta_{k}$ :
\begin{equation}
  \frac{1}{3} \, \sum_{k=1}^{3} <L_{k}f_{i}(r,\varphi), L_{k}f_{j}(r,\varphi)> \;
 = \frac{1}{3} \,  \frac{1}{\pi(1 - \rho_{1}^{2})}
      \; \sum_{k=1}^{3}\int {\rm d}\varphi
               \int_{\rho_{1}}^{1} \rho\, {\rm d}\rho\,
            (L_{k}f_{i}(\rho,\varphi)) \, (L_{k}f_{j}(\rho,\varphi))
    \label{eq:vecProdTiltDispl}
\end{equation}
After introducing equation~(\ref{eq:fTofn}), integrating over the azimuth angle $\varphi$,
and summing over the three directions denoted by $k$ one gets
\begin{eqnarray}
  \frac{1}{3} \, \sum_{k=1}^{3} <L_{k}f_{i}(r,\varphi), L_{k}f_{j}(r,\varphi)>
    & = & \frac{1}{8R^{4}} \, \frac{1 + \delta_{n,0}}{1 - \rho_{1}^{2}}
            \int_{\rho_{1}}^{1} \rho \, {\rm d}\rho \,
        \left[ (L_{n,1}f_{n,i}(\rho))(L_{n,1}f_{n,j}(\rho)) \, + \right.
         \nonumber \\
    & & \vspace{80mm} \left.  4\, (L_{n,2}f_{n,i}(\rho))(L_{n,2}f_{n,j}(\rho)) \right]
       \label{eq:vecProdTiltDispl1}
\end{eqnarray}
A straightforward calculation shows that
\begin{equation}
  L^{\rm F}\, L = \frac{1}{8R^{4}} \, L^{2}_{{\rm B},n}
   \label{eq:LFTiltDispl}
\end{equation}
The boundary functional is given by
\begin{eqnarray}
   \underline{\beta}_{{\rm L}}[f_{n,i}f_{n,j}] & = & 
      \frac{1}{8R^{4}} \,\frac{1+\delta_{n,0}}{1-\rho_{1}^{2}} \, \left\{
  \left [\left( \rho\frac{{\rm d}^{2}}{{\rm d}\rho^{2}}f_{n,i}(\rho)
       - \frac{{\rm d}}{{\rm d}\rho}f_{n,i}(\rho)
       + \frac{n^{2}}{\rho}f_{n,i}(\rho) \right) \,
             \frac{{\rm d}}{{\rm d}\rho}f_{n,j}(\rho) \right ]_{\rho_{1}}^{1} +
      \right.
            \nonumber\\
   & & \left.
     \left [\left( -\rho\frac{{\rm d}^{3}}{{\rm d}\rho^{3}}f_{n,i}(\rho)
       - \frac{{\rm d^{2}}}{{\rm d}\rho^{2}}f_{n,i}(\rho)
       + (1+3n^{2})\frac{1}{\rho}\frac{{\rm d}}{{\rm d}\rho}f_{n,i}(\rho)
          - 4\frac{n^{2}}{\rho^{2}}f_{n,i}(\rho) \right) \,
                                     f_{n,j}(\rho) \right ]_{\rho_{1}}^{1} \right\}
            \label{eq:boundaryFunctionalPTT}
\end{eqnarray}
With $\sigma$ related to ${\tilde \sigma}$ in equation~(\ref{eq:basicDiffEq}) by
$\sigma = \sqrt{8}R^{2} {\tilde \sigma}$ the fourth-order
differential operator appearing in
\begin{equation}
  L_{{\rm B},n}^{2} f_{n}(\rho) = \sigma^{2} f_{n}(\rho)
    \label{eq:4thorderDiffEq}
\end{equation}
can be written as a product of two second-order differential operators.
Equation~(\ref{eq:4thorderDiffEq}) then becomes
\begin{equation}
  (L_{{\rm B},n} + \sigma)\, (L_{{\rm B},n} - \sigma) f_{n}(\rho) = 0
   \label{eq:PTTDiffEqFact}
\end{equation}
With the same definition (\ref{eq:defLambda}) of $\lambda$ as in
section~\ref{sec:1DimTilt}
the general solution of (\ref{eq:PTTDiffEqFact}) is a sum of the Bessel functions
$J_{n}(\lambda \rho)$, $Y_{n}(\lambda \rho)$, $I_{n}(\lambda \rho)$,
and $K_{n}(\lambda \rho)$ with coefficients $c_{{\rm J}}$, $c_{{\rm Y}}$,
$c_{{\rm Y}}$ and $c_{{\rm K}}$ :
\begin{equation}
  f(\rho) = c_{{\rm J}}J_{n}(\lambda\rho) +
            c_{{\rm Y}}Y_{n}(\lambda\rho) +
            c_{{\rm I}}I_{n}(\lambda\rho) +
            c_{{\rm K}}K_{n}(\lambda\rho)
   \label{eq:solutionsBesselPTTHole}
\end{equation}
For an arbitrary choice of the functions $f_{n,j}$ the boundary
functional~(\ref{eq:boundaryFunctionalPTT}) is zero
if the expressions in both square
brackets vanish for $\rho=\rho_{1}$ and $\rho=1$.
Introducing (\ref{eq:solutionsBesselPTTHole}) into
the four boundary conditions
gives a system of four linear homogeneous equations in
$c_{{\rm J}}$, $c_{{\rm Y}}$, $c_{{\rm I}}$, and $c_{{\rm K}}$ from which the
infinite set of normalised wavenumbers $\lambda_{n,i}$ and the corresponding ratios
of $c_{{\rm Y}}$, $c_{{\rm I}}$, and $c_{{\rm K}}$ to $c_{{\rm J}}$
can be obtained.
Similarly to the corresponding one-dimensional example the singular values $\sigma_{n,i}$
are the squares of the wavenumbers $\lambda_{n,i}$.

\begin{figure}[h]
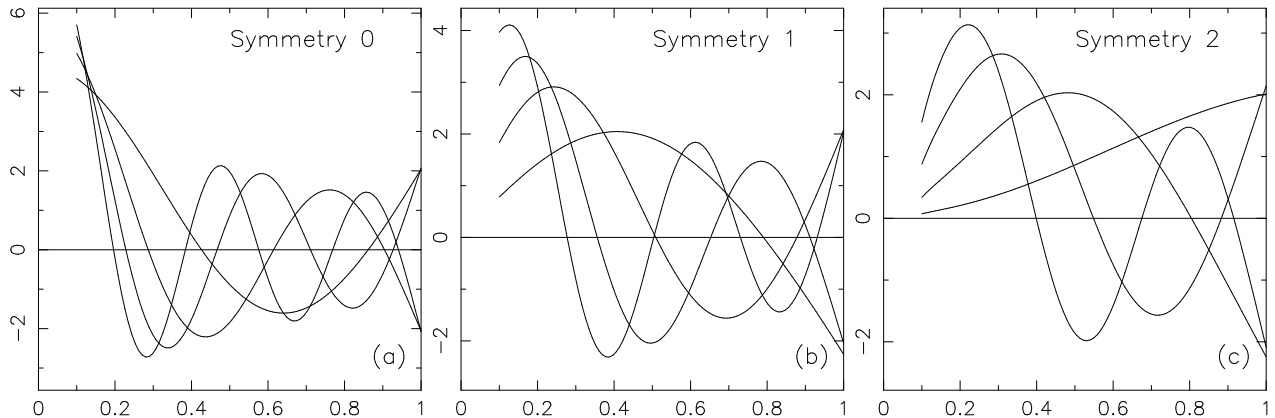

     \centerline{\hbox{
      \psfig{figure=figure7a.ps,width=55mm}
      \psfig{figure=figure7b.ps,width=55mm}
      \psfig{figure=figure7c.ps,width=55mm}}}
   \caption{\label{fig:figure7} {\small Lowest analytical
    normal modes of an annular mirror with $\rho_{1} = 0.1$ for piston,
    tip and tilt movements for the rotational symmetries 0, 1, and 2.}}
\end{figure}

For the rotational symmetries 0, 1, and 2 and $\rho_{1} = 0.1$
figure~\ref{fig:figure7}
shows the four analytical normal modes with the lowest singular values $\sigma_{n,i}$.
Using the classification of the modes introduced in
section~\ref{sec:2DimAnalyticalP} three modes are missing compared with the set of
Zernike modes. Two of them are the two
lowest modes of rotational symmetry 0 without and with one node, corresponding to
Zernike piston and defocus,
and the third one is the lowest mode of
rotational symmetry 1 corresponding to Zernike tilt.
All three modes can be generated without relative vertical
displacements at intersegment edges.

The functions are somewhat similar to the corresponding Zernike polynomials, but should
better be compared with the elastic minimum energy modes of circular plates
(see section~\ref{sec:compElastic}).
For the case of a full mirror, that is $\rho_{1}=0$, the Bessel functions 
$Y_{n}(\lambda \rho)$ and $K_{n}(\lambda \rho)$, which diverge at $\rho = 0$, drop out
of the general solution, which is then given by
\begin{equation}
  f = c_{{\rm J}}J_{n}(\lambda\rho) +
      c_{{\rm I}}I_{n}(\lambda\rho)
   \label{eq:solutionsBesselPTTNoHole}
\end{equation}
The expressions in the two square brackets of the boundary
functional~(\ref{eq:boundaryFunctionalPTT}) are
automatically zero for $\rho=0$. This leaves two boundary conditions which are
sufficient to calculate the wavenumbers $\lambda_{n,i}$ and the corresponding ratios
$(c_{{\rm I}}/c_{{\rm J}})_{n,i}$. The functions look similar to the ones in
figure~\ref{fig:figure7}, but now, for all symmetries $n>0$, they
are zero at the centre of the mirror.
\section{Comparison with SVD results}
\label{sec:SVD2Dim}
\subsection{SVD programs}
\label{sec:compSvd2Dim}
\begin{figure}[h]
     \centerline{\hbox{
      \psfig{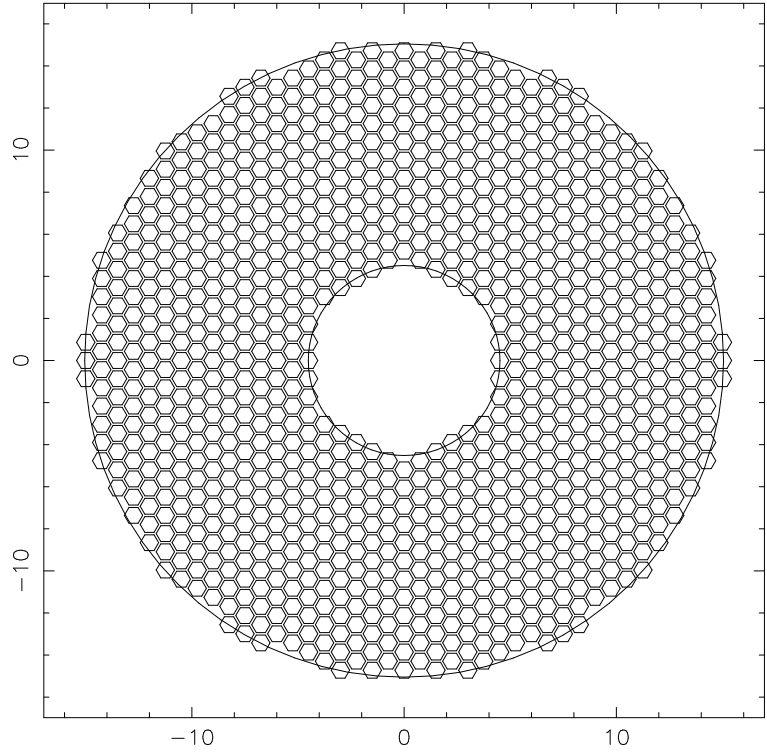}}}
   \caption{\label{fig:figure8} {\small Segmentation used for the
             SVD calculations.}}
\end{figure}

All SVD calculations were done with computer programs supplied by G. Chanan [5]. They were
developed
for and applied to the control of the segmented primary mirror of the Keck telescope.
Most of the data used in this paper were obtained for the segmented mirror shown in
figure~\ref{fig:figure8}.
In the following, the radius of the segments is, as shown in
figure~\ref{fig:figure5}, denoted by $a$ and the distance between the sensors
along one intersegment edge by $b$.
For the numerical SVD-calculations we used the parameters $a=0.5\,$m, $b=0.3078\,$m,
and $R=15.05\,$m.
The segments in figure~\ref{fig:figure8} are the ones
with the centres inside the range of normalised
radii $\rho$ of $0.30 < \rho < 1.0$. This mirror has got 1002 segments and 5688 sensors.
The radius of the circle $r_{{\rm act}}$ defined by the actuator locations
shown in figure~\ref{fig:figure5} was $r_{{\rm act}}=0.52275\,a$.
Calculations were also done for a geometry similar to the one shown in
figure~\ref{fig:figure8}, but without a central hole. This mirror has
got 1099 segments and 6348 sensors.
\subsection{Piston movements}
\subsubsection{Mode shapes}
\begin{figure}[h]
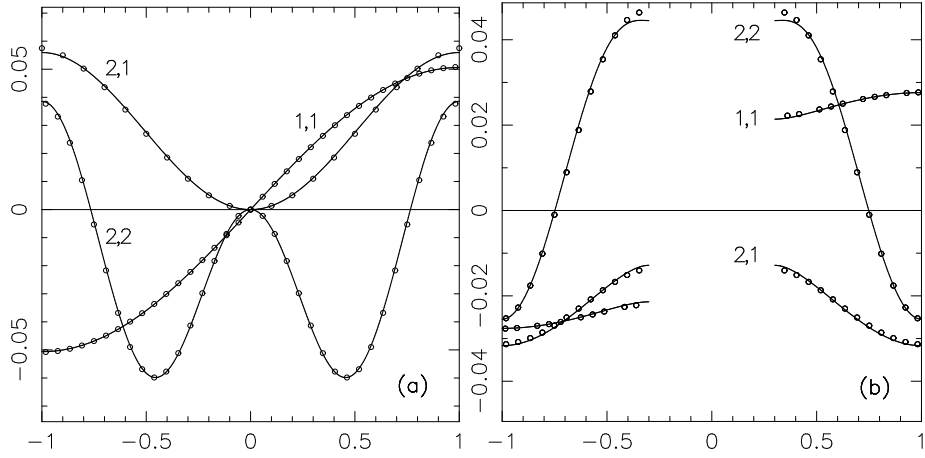

     \centerline{\hbox{
      \psfig{figure=figure9a.ps,width=60mm}
      \psfig{figure=figure9b.ps,width=60mm}}}
   \caption{\label{fig:figure9} {\small Normal modes calculated by SVD (circles)
         and analytical normal modes (solid lines) for pure piston movements.
         The first index along a curve denotes the rotational symmetry and
         the second one the order within the symmetry.
         {\bf (a) :} Full mirror, {\bf (b) :} Annular mirror with $\rho_{1} = 0.3$.}}
\end{figure}
The number of SVD modes with non-zero singular values is $n_{{\rm nz}} = n_{{\rm seg}} - 1$.
One mode with a singular value equal to zero, which corresponds to the missing analytical mode
in section~\ref{sec:2DimAnalyticalP}, is related to a uniform piston movement of all segments.
For a few low order modes figure~\ref{fig:figure9}
shows the traces along a diameter for a full and an annular mirror.
Especially for the full mirror the agreement between the
shapes of the normal modes obtained with SVD (circles) and analytically (solid lines)
in section~\ref{sec:2DimAnalyticalP} is excellent. The deviations close to the inner hole
of the annular mirror are to be expected since this edge is less accurately defined
because of the smaller number of segments.
\subsubsection{Singular values}
\label{sec:singularValuesP}
For pure piston movements a sensor signal is proportional to the difference
between the values of an analytical normal mode $f_{n,i}$ at adjacent segment centres.
To the first approximation, this is proportional to the slopes of $f_{n,i}$ midway between
two centres along the line connecting the two centres 
times the distance $\sqrt{3}a$ between the two centres.
The mean square $\epsilon_{{\rm sig},n,i}^{2}$ of the sensor signals of the mode $(n,i)$
is therefore
\begin{equation}
  \epsilon^{2}_{{\rm sig},n,i} = \left( \sqrt{3}a \right)^{2} \;
                         \frac{1}{n_{{\rm sig}}} \;
                          \sum_{j=1}^{n_{{\rm sig}}} T^{2}_{n,i,j} \; ,
    \label{eq:pistonSumSignal}
\end{equation}
where $T_{n,i,j}$ is the slope of the mode $i$ of rotational symmetry $n$
perpendicular to the edge at the location of the sensor $j$.
For a very large number of segments the averaged sum in equation~(\ref{eq:pistonSumSignal})
can be replaced by the integral expressions in equation~(\ref{eq:vecProdPiston1}) with $i=j$ :
\begin{equation}
  \epsilon^{2}_{{\rm sig,an},n,i} = 
                    \frac{3 a^{2}}{2 R^{2}} \;
                   \frac{1 + \delta_{n,0}}{1 - \rho_{1}^{2}}
              \, \int_{\rho_{1}}^{1} \rho\, {\rm d}\rho \,
              \left[ \left( \frac{{\rm d}}{{\rm d} \rho}f_{n,i}(\rho) \right)^{2} +
                       \frac{n^{2}}{\rho^{2}}f^{2}_{n,i}(\rho) \right]
         \label{eq:rmsSensorsP}
\end{equation}
After dividing by the mean square $\epsilon^{2}_{{\rm act,an},n,i}$
of the actuator positions
\begin{equation}
  \epsilon^{2}_{{\rm act},n,i} = \frac{1 + \delta_{n,0}}{1 - \rho_{1}^{2}} \,
            \int_{\rho_{1}}^{1} \rho\, {\rm d}\rho \, f_{n,i}^{2}(\rho)
   \label{eq:rmsF}
\end{equation}
one gets
\begin{equation}
{\displaystyle
  \frac{\epsilon^{2}_{{\rm sig,an},n,i}}{\epsilon^{2}_{{\rm act,an},n,i}} =
          \frac{3 a^{2}}{2R^{2}} \; \sigma_{n,i}^{2}
   = 3 a^{2} \,  {\tilde \sigma}_{n,i}^{2}
    \label{eq:singularValueSVDP}
}
\end{equation}
where $\sigma_{n,i} = \sqrt{2}R {\tilde \sigma}_{n,i}$ are the normalised analytical
singular values.
Similarly to section~\ref{sec:comp1DimExa} one finally gets
for the relationship between a singular value $\sigma_{{\rm svd,an},n,i}$
expected analytically from SVD and the corresponding $\sigma_{n,i}$
\begin{equation}
  \sigma_{{\rm svd,an},n,i} = \sqrt{\frac{3}{2}\frac{n_{{\rm sig}}}{n_{{\rm act}}}}
                    \, \frac{a}{R} \; \sigma_{n,i}
     \label{eq:singularValueSVDAnalytical}
\end{equation}
Since the ratio $n_{{\rm sig}}/n_{{\rm act}}$ is effectively constant for highly segmented
mirrors, the singular values for a specific mode $(n,i)$ expected from SVD scale with $a/R$.
\begin{figure}[h]
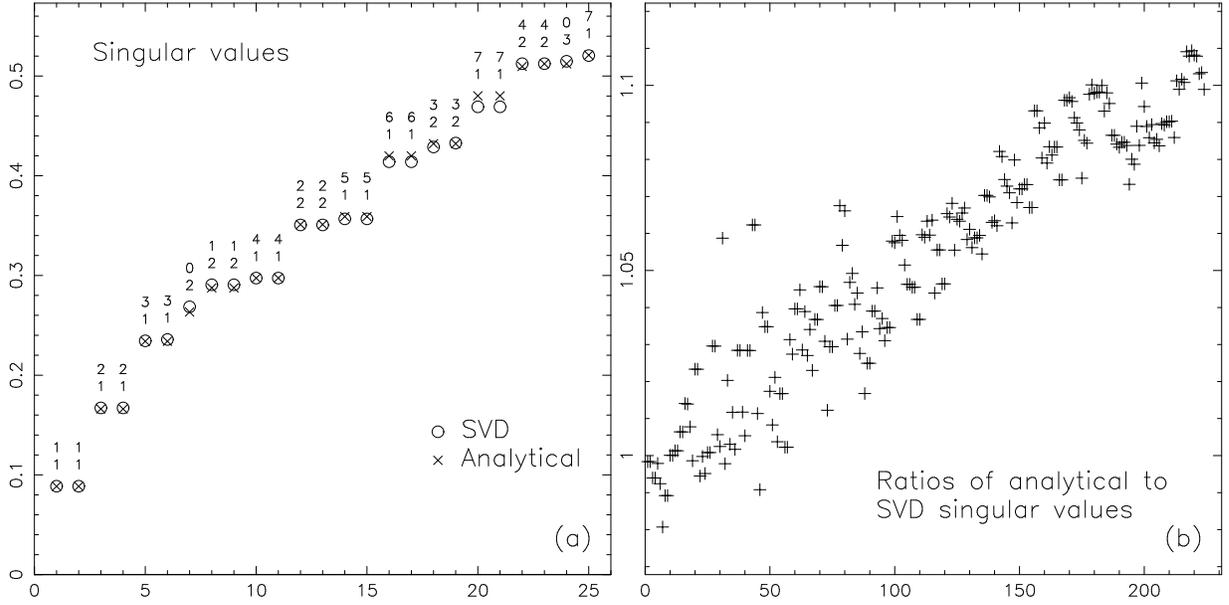

     \centerline{\hbox{
      \psfig{figure=figure10a.ps,width=80mm}
      \psfig{figure=figure10b.ps,width=80mm}}}
   \caption{\label{fig:figure10} {\small Singular values calculated
     by SVD (circles) and analytically (crosses) for pure piston movements.
     {\bf (a) :} Lowest 25 modes,
     {\bf (b) :} Ratios of analytically expected singular values to the ones
     calculated by SVD for the lowest 230 modes.}}
\end{figure}
Figure~\ref{fig:figure10}a shows for the lowest 25 normal modes
the true singular values $\sigma_{{\rm svd},n,i}$ given by SVD (circles) on the one hand
and the ones $\sigma_{{\rm svd,an},n,i}$
(crosses) according to equation~(\ref{eq:singularValueSVDAnalytical}) on the other hand.
The agreement is quite satisfactory. Figure~\ref{fig:figure10}b,
displaying the ratios of $\sigma_{{\rm svd,an},n,i}$ to $\sigma_{{\rm svd},n,i}$, shows
that for the higher modes the analytical calculation overstimates the singular values.

\subsubsection{Error propagation}
Starting from equation~(\ref{eq:singularValueSVDP}) one can derive how
the total r.m.s. $\epsilon_{{\rm act,n,an}}$ of the actuator errors generated by a given r.m.s.
$\epsilon_{{\rm sig,n}}$ of the signal errors scales with the number of segments.
The procedure is the same as the one used in section~\ref{sec:comp1DimExa}. One gets
\begin{equation}
  \epsilon_{{\rm act,n,an}}^{2} = \epsilon_{{\rm sig,n}}^{2}  \,
               \frac{2R^{2}}{3a^{2}} \, \frac{1}{n_{{\rm sig}}} \,
            \sum_{n,i}{\frac{1}{\sigma_{n,i}^{2}}},
     \label{eq:epsilonActPiston}
\end{equation}
where the sum runs over the $n_{{\rm nz}}=n_{{\rm seg}} - 1$ non-zero analytical singular values.
For a large number of segments the total number of segments is approximately given by
\begin{equation}
  n_{{\rm seg}} \approx \frac{2\pi}{3\sqrt{3}} \, \left(\frac{R}{a} \right)^{2}
   \label{eq:numberSegments}
\end{equation}
Neglecting the missing sensors at the edge of the mirror, the number of sensors is
approximately given by
\begin{equation}
      n_{{\rm sig}} \approx 6 \, n_{{\rm seg}}
      \label{eq:numSignals}
\end{equation}
Introducing all this into
equation~(\ref{eq:epsilonActPiston}) one gets
\begin{equation}
  \epsilon_{{\rm act,n,an}}^{2} \approx \epsilon_{{\rm sig,n}}^{2} \,
         \frac{1}{2\pi\sqrt{3}} 
           \sum_{n,i}{\frac{1}{\sigma_{n,i}^{2}}} 
	    \label{eq:rmsActuatorsPiston1}
\end{equation}
For the higher modes the analytical singular values increase approximately linearly both
with the symmetry $n$ and the order $i$, in the latter case, as one should expect,
in steps of $\pi$.
With such linear dependencies on $n$ and $i$ the sum in equation~(\ref{eq:rmsActuatorsPiston1})
does not converge if the number of segments goes to infinity. But it increases only very slowly,
namely approximately linearly with the logarithm of $n_{{\rm nz}}$.
Therefore, $\epsilon_{{\rm act,n,an}}$ scales approximately with $log(R/a)$.

For the full mirror desribed in section~\ref{sec:compSvd2Dim} the sum in
equation~(\ref{eq:rmsActuatorsPiston1}) is approximately
equal to 2.4. This gives for the r.m.s. $\epsilon_{{\rm act,n,an}}$ of the actuator errors
\begin{equation}
  \epsilon_{{\rm act,n,an}} \approx 0.470 \; \epsilon_{{\rm sig,n}}
	    \label{eq:rmsActuatorsPiston}
\end{equation}
This is in good agreement with the true r.m.s.
$\epsilon_{{\rm act,n,svd}} = 0.504 \, \epsilon_{{\rm sig,n}}$
of the actuator errors calculated with the true singular values
$\sigma_{{\rm svd},n,i}$. Similar values for $\epsilon_{{\rm act,n,svd}}$
are given in table 4 of reference [6].
The true value $\epsilon_{{\rm act,n,svd}}$ is slightly larger than $\epsilon_{{\rm act,n,an}}$
since, first, the singular
values $\sigma_{{\rm svd,an},n,i}$ are larger than the true ones $\sigma_{{\rm svd},n,i}$
and, second, the number of signals is smaller than the value of $6 \, n_{{\rm seg}}$
used in the derivation of equation~(\ref{eq:rmsActuatorsPiston}).
\subsection{Piston, tip and tilt movements}
\subsubsection{Mode shapes}
\begin{figure}[h]
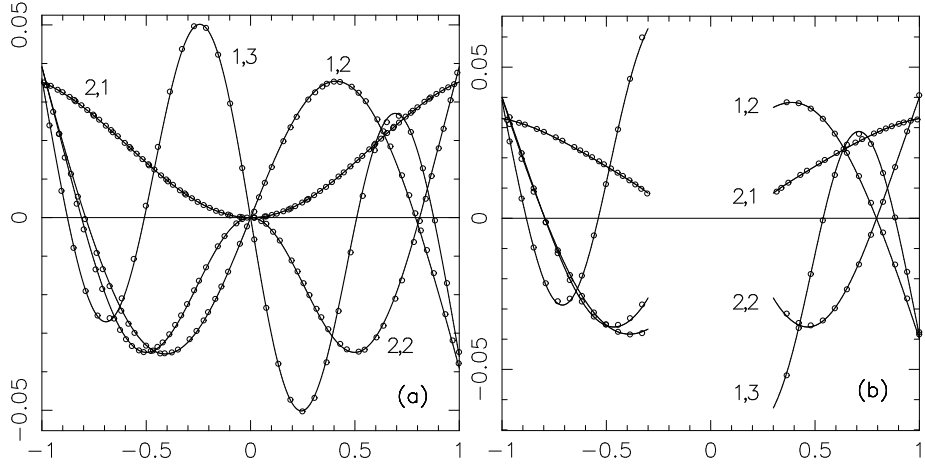

     \centerline{\hbox{
      \psfig{figure=figure11a.ps,width=60mm}
      \psfig{figure=figure11b.ps,width=60mm}}}
   \caption{\label{fig:figure11} {\small Normal modes calculated by SVD (circles)
         and analytical normal modes (solid lines) for piston, tip and tilt movements
         for the lowest two modes of the rotational symmetries
         1 and 2. {\bf (a) :} Full mirror,
         {\bf (b) :} Annular mirror with $\rho_1 = 0.3$.}}
\end{figure}

The number of non-zero singular values is $n_{{\rm nz}} = n_{{\rm act}} - 4$.
Four normal modes calculated by SVD with singular values zero are the piston mode,
the defocus mode and the two tilt modes.
They correspond to the three missing analytical normal modes in section~\ref{sec:derivModesPTT}.
Figure~\ref{fig:figure11} shows for a few of the lowest modes
the SVD solutions (circles) and the corresponding analytical solutions (solid lines)
for a full and an annular mirror.
Especially for the full mirror the agreement is excellent.
\subsubsection{Singular values}
\label{sec:SingularValuesPTT}
The sensor signals are proportional to the difference in the slope of an analytical
normal mode $f_{n,i}$ between adjacent segment centres.
If only the torsion contributes to a sensor signal one has to take the slope
perpendicular to the line connecting the two centres. The difference in the slope is then,
to the first approximation, proportional to the torsion of $f_{n,i}$ at the location midway
between the two centres, the distance $\sqrt{3}a$ between the two centres,
and the distance from the midway point to the sensor, that is $b/2$.
The mean square $\epsilon_{{\rm sig},n,i}^{2}$
of the sensor signals for the mode $(n,i)$ is therefore
\begin{equation}
  \epsilon^{2}_{{\rm sig},n,i} = \left(\frac{\sqrt{3}ab}{2} \right)^{2} \;
                         \frac{1}{n_{{\rm sig}}} \;
                          \sum_{j=1}^{n_{{\rm sig}}} T^{2}_{n,i,j} \; ,
     \label{eq:PTTSumSignals}
\end{equation}
where $T_{n,i,j}$ is the torsion of the mode $(n,i)$ at the location of the sensor $j$.
For a very large number of segments the averaged sum in equation~(\ref{eq:PTTSumSignals})
can be replaced by the integral expressions in
equation~(\ref{eq:vecProdTiltDispl1}) with $i=j$ :
\begin{equation}
  \epsilon^{2}_{{\rm sig,an},n,i} = 
                \frac{3 a^{2}b^{2}}{32R^{4}} \;
                   \frac{1 + \delta_{n,0}}{1 - \rho_{1}^{2}}
              \, \int_{0}^{1} \rho{\rm d} \, \rho \,
                   [(L_{n,1}f_{n,i}(\rho))^{2} + 4(L_{n,2}f_{n,i}(\rho))^{2}]
         \label{eq:rmsSensorsPTT}
\end{equation}
After dividing, as in
section~\ref{sec:singularValuesP}, by the mean square of the actuator positions,
the ratio of the mean square of the sensors signals to the mean square of the
actuator positions for a given mode $(n,i)$ becomes
\begin{equation}
  \frac{\epsilon^{2}_{{\rm sig,an},n,i}}{\epsilon^{2}_{{\rm act,an},n,i}} =
          \frac{3 a^{2}b^{2}}{32R^{4}} \; \sigma_{n,i}^{2}
      = \frac{3 a^{2}b^{2}}{4} \; {\tilde \sigma}_{n,i}^{2} \, ,
    \label{eq:singularValueSVDPTT}
\end{equation}
where $\sigma_{n,i} = \sqrt{8}R^{2}{\tilde \sigma}_{n,i}$ are the normalised
analytical singular values. Similarly to section~\ref{sec:1DimPiston}
one finally gets
for the relationship between the singular value $\sigma_{{\rm svd,an},n,i}$
of the mode $(n,i)$ expected analytically from SVD and the corresponding normalised
analytical one $\sigma_{n,i}$
\begin{equation}
  \sigma_{{\rm svd,an},n,i} = \sqrt{\frac{3}{32}\frac{n_{{\rm sig}}}{n_{{\rm act}}}}
                    \, \frac{ab}{R^{2}} \; \sigma_{n,i}
     \label{eq:singularValueSVDAnalyticalPTT}
\end{equation}
Since the ratio $n_{{\rm sig}}/n_{{\rm act}}$ is effectively constant for highly segmented
mirrors, the singular values expected from SVD scale with $ab/R^{2}$.
\begin{figure}[h]
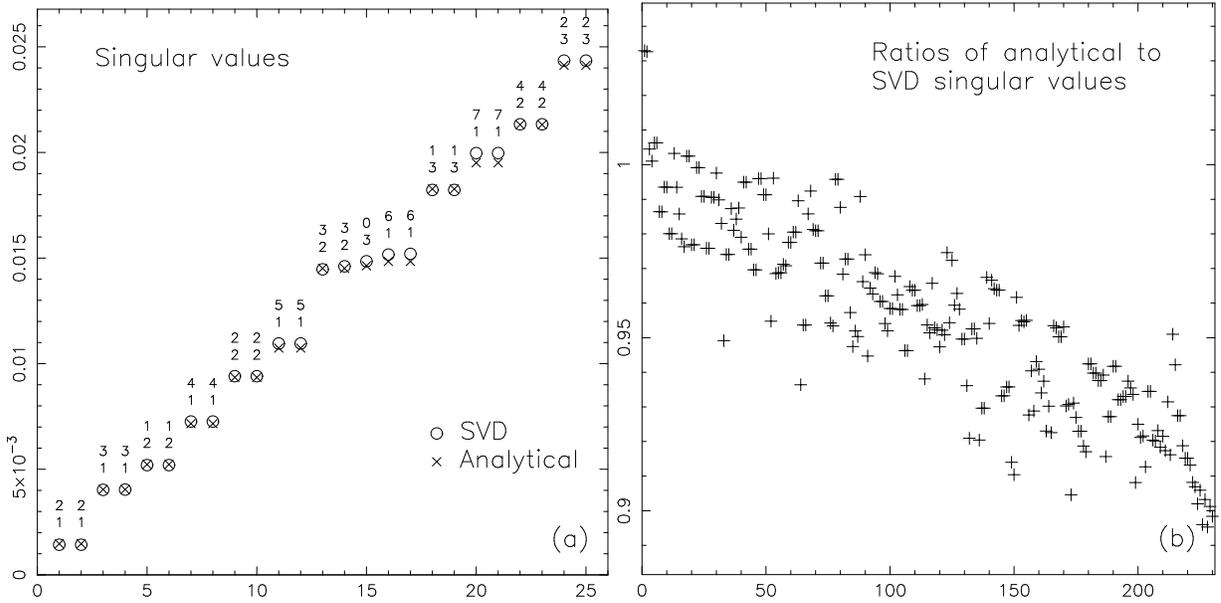

     \centerline{\hbox{
      \psfig{figure=figure12a.ps,width=80mm}
      \psfig{figure=figure12b.ps,width=80mm}}}
   \caption{\label{fig:figure12} {\small Singular values calculated with SVD
     (circles) and analytically (crosses) for piston, tip and tilt movements.
     {\bf (a)} : Lowest 25 modes,
     {\bf (b)} : Ratios of expected singular values to the ones calculated by SVD
     for the lowest 230 modes.}}
\end{figure}
Figure~\ref{fig:figure12}a shows for the lowest 25 normal modes the
true singular values $\sigma_{{\rm svd},n,i}$
given by SVD (circles) on the one hand and the ones $\sigma_{{\rm svd,an},n,i}$ (crosses)
according to equation~(\ref{eq:singularValueSVDAnalyticalPTT}) on the other hand.
The agreement is quite satisfactory. Figure~\ref{fig:figure12}b,
displaying the ratios of $\sigma_{{\rm svd,an},n,i}$ to $\sigma_{{\rm svd},n,i}$, shows
that for the higher modes the analytical calculation understimates the singular values.
\subsubsection{Error propagation}
Starting from equation~(\ref{eq:singularValueSVDPTT}) one can derive an expression
for the total r.m.s. $\epsilon_{{\rm act,n,an}}$ of the actuator errors based on
the analytical singular values.
The procedure is again the same as the one used 
in section~\ref{sec:comp1DimExa}. One gets
\begin{equation}
  \epsilon_{{\rm act,n,an}}^{2} = \epsilon_{{\rm sig,n}}^{2} \,
        \frac{32R^{4}}{3a^{2}b^{2}} \, \frac{1}{n_{{\rm sig}}} \,
            \sum_{n,i}{\frac{1}{\sigma_{n,i}^{2}}},
	    \label{eq:rmsActuatorsPTT1}
\end{equation}
where the sum runs over the $n_{{\rm nz}}= n_{{\rm act}}-4$ non-zero analytical singular values.
For the higher modes the square roots of the analytical singular values 
increase approximately linearly both with the symmetry $n$ and the order $i$,
in the latter case, as one should expect, in steps of $\pi$.
The sum in equation~(\ref{eq:rmsActuatorsPTT1}) converges rapidly for a large number
of segments.
For the full mirror described in section~\ref{sec:compSvd2Dim} the limit value
is approximately 0.065.
Introducing this and the approximate expressions
(\ref{eq:numberSegments}) and (\ref{eq:numSignals}) into
equation~(\ref{eq:rmsActuatorsPTT1}) one finally gets
for the r.m.s. of the actuator errors
\begin{equation}
  \epsilon_{{\rm act,n,an}} \approx 0.31 \, \frac{R}{b} \; \epsilon_{{\rm sig,n}}
	    \label{eq:rmsActuatorsPTT}
\end{equation}
The true r.m.s. $\sigma_{{\rm act,n,svd}}$ of the
actuator errors is effectively the same as the one calculated from
equation~(\ref{eq:rmsActuatorsPTT}).
For the currently envisaged extremely large telescopes the ratio $R/b$ is approximately
equal to 50.
The r.m.s. of the actuator errors is then approximately fifteen times larger than the r.m.s.
of the signal errors. This is in good agreement with the results of SVD calculations for
a thirty meter telescope given in table 1 of reference [6].
\subsection{Piston, tip and tilt generated by relative tilt signals}
\label{sec:relativeTilt}
Some sensors detect in addition to the relative vertical displacements also the relative
tilts between adjacent segments.
One could use the same formalism as in the other examples to derive the differential equations
and the boundary conditions. Unfortunately, the fourth-order differential operator cannot
easily be factorised into two second-order differential operators as it could be done in
equations~(\ref{eq:4thorderDiffEq}) and (\ref{eq:PTTDiffEqFact}).
But usually the signals due to the relative tilts perpendicular to the edges
are much smaller than the signals due to
relative vertical displacements [6]. Most modes will then hardly be affected by the
tilt signals. There is only one additional mode of rotational symmetry zero with
a non-zero singular value which converges, in the limit of the signals due to relative
tilts going to zero, to the pure defocus mode with a singular value equal to zero.
For small, but still non-zero tilt signals the singular value of this mode can be estimated.
It is related to the integral of the mean
curvature over the mirror for a mode shape proportional to pure defocus.
The operator which measures the curvature along a direction $\theta_{k}$ is given by
\begin{eqnarray}
   L_{k}f(r,\varphi) = 
  \frac{\partial^{2} f(r,\varphi)}{\partial^{2} s_{\theta_{k}}}
   & = &      \frac{1}{R^{2}} \; \left\{ \cos(n\varphi)
           \left[ \cos^{2}(\theta_{k})\frac{\partial^{2}f_{n}(\rho)}{\partial \rho^{2}}
                + \sin^{2}(\theta_{k})\frac{1}{\rho}
                   \left( \frac{\partial f_{n}(\rho)}{\partial \rho}
                    - \frac{n^{2}}{\rho}\,f_{n}(\rho) \right) \right] \right.
                        \nonumber \\
   &  &  \left.   + \sin(n\varphi)\, \sin(2\theta_{k})\, n \,
                \left( \frac{\partial f_{n}(\rho)}{\partial \rho}
                    - \frac{1}{\rho}\,f_{n}(\rho) \right) \right\}
    \label{eq:operAvgCurv}
\end{eqnarray}
One then obtains,
after introducing equation~(\ref{eq:fTofn}), integrating over the azimuth angle $\varphi$,
and averaging over the three directions denoted by $k$,
for the special case of rotational symmetry 0 and a full mirror
\begin{equation}
  \frac{1}{3} \, \sum_{k=1}^{3} <L_{k}f_{i}(r,\varphi), L_{k}f_{i}(r,\varphi)>
         = \frac{1}{4R^{4}}\,
         \int_{0}^{1} \rho \, {\rm d}\rho \,
         \left[ 3 \left( \frac{{\rm d}^{2}f_{0,i}(\rho)}{{\rm d}\rho^{2}} \right)^{2}
        + \frac{3}{\rho^{2}} \left( \frac{{\rm d}f_{0,i}(\rho)}{{\rm d}\rho} \right)^{2}
        + \frac{2}{\rho}\, \frac{{\rm d}^{2}f_{0,i}(\rho)}{{\rm d}\rho^{2}}
           \frac{{\rm d}f_{0,i}(\rho)}{{\rm d}\rho} \right]
\end{equation}
If one assumes that the lowest mode $f_{0,2}$
($f_{0,1}$ would denote the piston mode) is equal to pure Zernike defocus 
$c\sqrt{3}(2\rho^{2} - 1)$, where $c$ is the r.m.s. of this mode, one gets
\begin{equation}
  \frac{1}{3} \, \sum_{k=1}^{3} <L_{k}f_{0,2}(r,\varphi), L_{k}f_{0,2}(r,\varphi)>
       = c^{2}\, \frac{48}{R^{4}}
  \label{eq:avgCurvEnergyDefocus}
\end{equation}
The sensor signal is proportional to the difference in tilt between adjacent segment centres.
If only the curvature contributes to the sensor signal one has to take the tilt component
parallel to the line connecting the two centres. The difference in tilt is then,
to the first approximation, proportional to the curvature at the location midway between the
two centres, the distance $\sqrt{3}a$ between the centres and the lever $g$ which converts the
angular difference into a differential displacement [6].
The mean square $\epsilon_{{\rm sig},0,2}^{2}$ of the sensor signals is therefore
\begin{equation}
  \epsilon^{2}_{{\rm sig},0,2} = (\sqrt{3}\,ag)^{2} \;
                         \frac{1}{n_{{\rm sig}}} \;
                          \sum_{j=1}^{n_{{\rm sig}}} T^{2}_{0,2,j} \; ,
     \label{eq:TiltSumSignals}
\end{equation}
where $T_{0,2,j}$ is the curvature of the defocus mode at the location of the sensor $j$.
In the limit of small segments the averaged sum in equation~(\ref{eq:TiltSumSignals})
can be replaced by the integral expression in equation~(\ref{eq:avgCurvEnergyDefocus}) :
\begin{eqnarray}
  \epsilon^{2}_{{\rm sig,an},0,2} & = & 144\;  c^{2} \, \left( \frac{ag}{R^{2}} \right)^{2}
                      \label{eq:rmsSensorsTilt}
\end{eqnarray}
After dividing, as in
section~\ref{sec:singularValuesP}, by the mean square $c^{2}$ of the position function,
the ratio of the r.m.s. of the sensor signals to the r.m.s. of the
actuator positions for the defocus mode becomes
\begin{equation}
{\displaystyle
  \frac{\epsilon_{{\rm sig,an},0,2}}{\epsilon_{{\rm act,an},0,2}} =
          12 \, \frac{ag}{R^{2}} }
    \label{eq:singularValueTilt}
\end{equation}
The singular value expected from SVD is then finally given by
\begin{equation}
  \sigma_{{\rm svd,an},0,2} = 12 \, \sqrt{\frac{n_{{\rm sig}}}{n_{{\rm act}}}} \,
               \frac{ag}{R^{2}}
    \label{eq:singularValueSVDTilt}
\end{equation}
For the full mirror described in section~\ref{sec:compSvd2Dim} with $g=23.53\,$mm
one gets from equation~(\ref{eq:singularValueSVDTilt})
$\sigma_{{\rm svd,an},0,2} = 0.0007206$. This is in good agreement with the result
of $\sigma_{{\rm svd},0,2} = 0.0007064$ obtained from a SVD calculation.
\section{Relationship to modes minimizing the signals measured by the sensors}
\label{sec:minRmsSignal}
For arbitrary position functions the mean square ${\cal J}$ of the signals is given 
$f(\xi)$ by
\begin{equation}
   {\cal J} = \frac{1}{2D} \, \int (Lf(\xi))^{2}\, {\rm d}\xi
\end{equation}
for the two one-dimensional examples with the operators $L$ defined in
equations~(\ref{eq:diffOperExa}) and (\ref{eq:diffOperExb}), and for arbitrary
position functions $f_{n}(\rho)$ by
\begin{eqnarray}
  {\cal J} & = & \frac{1}{2R^{2}} \, \frac{1 + \delta_{n,0}}{1 - \rho_{1}^{2}} \,
            \int_{\rho_{1}}^{1} \rho \, {\rm d}\rho \, \left(
        (\frac{{\rm d}}{{\rm d}\rho}f_{n}(\rho))^{2} +
                     \frac{n^{2}}{\rho^{2}}f^{2}_{n}(\rho) \right)
                 \label{eq:energySignalsP} \\
  {\cal J} & = & \frac{1}{8R^{4}} \, \frac{1 + \delta_{n,0}}{1 - \rho_{1}^{2}} \,
                    \int_{\rho_{1}}^{1} \rho \, {\rm d}\rho \,
       \left( (L_{\rho,1}f_{n}(\rho))^{2} + 4(L_{\rho,2}f_{n}(\rho))^{2} \right)
                 \label{eq:energySignalsPTT}
\end{eqnarray}
for the two-dimensional examples in sections~\ref{sec:2DimAnalyticalP}
and \ref{sec:2DimAnalyticalPTT}.
On the assumption that the mean square ${\cal C}$ of the position functions,
defined by
\begin{eqnarray}
  {\cal C} & = & \frac{1}{2} \, \int_{-1}^{+1} \, f^{2}(\xi) \, {\rm d}\xi \\
  {\cal C} & = & \frac{1 + \delta_{n,0}}{1-\rho_{1}^{2}} \,
          \int_{\rho_{1}}^{1} \rho \, {\rm d}\rho \, f^{2}_{n}(\rho)
\end{eqnarray}
for the one- and two-dimensional cases, is constant,
the modes which minimise the r.m.s. of the signals can be calculated by
variational methods. Denoting the variation by $\delta$ one has to solve
 \begin{equation}
   \delta ({\cal J} - \zeta {\cal C}) = 0
\end{equation}
where $\zeta$ is a free parameter which can be interpreted the energy
per unit of the r.m.s. of ${\cal C}$.
This gives, with $m=0$ and $\eta=\xi$ for the one-dimensional
and $m=1$, $\eta=\rho$, and $f=f_{n}$ for the two-dimensional examples
\begin{equation}
   \frac{\delta J}{\delta f} - 2\zeta \eta^{m} f = 0
\end{equation}
or explicitly, with a prime denoting a partial derivative with respect to $\eta$,
\begin{equation}
   \frac{\partial J}{\partial f}
     - \frac{\partial}{\partial \eta}\frac{\partial J}{\partial f'}
     + \frac{\partial^{2}}{\partial \eta^{2}}\frac{\partial J}{\partial f''}
     - 2\zeta \eta^{m} f = 0
     \label{eq:explVar}
\end{equation}
For all examples given in this paper the differential equations derived
from equation~(\ref{eq:explVar}) are identical to the ones derived with the method
outlined in section~\ref{sec:introMethod}.
The analytical normal modes can therefore be regarded as a special class of the modes
which minimise the r.m.s. of the signals for a specified r.m.s. of the position
function $f$.
Their special characteristic  is that the
position modes and corresponding signal modes form orthogonal sets, which defines
the boundary conditions.
\section{Relationship of analytical normal modes to elastic modes}
\label{sec:compElastic}
\subsection{Reasons for a comparison}
In future extremely large telescopes segmented mirrors will be used together with large
flexible meniscus mirrors. Whereas the aberrations of the segmented mirrors are best
described by the analytical normal modes derived in this paper, the deformations of the thin
meniscus mirros are best described by minimum elastic energy modes [3]. For the two
examples in sects.~\ref{sec:1DimTilt} and \ref{sec:2DimAnalyticalPTT}
both types of modes are closely related and, as will be shown in this section, often
very similar. This offers the possibility to correct aberrations, which are generated
by segmented mirrors, by thin meniscus mirrors and vice versa.
\subsection{One-dimesional segmentation}
The differential equation (\ref{eq:diffEqExb1}) for example (b) in
section~\ref{sec:1DimTilt} is identical to the
one for minimum elastic energy modes of a bar. Since the boundary conditions
(\ref{eq:BC1Exb}) and (\ref{eq:BC2Exb}) are also the same as the boundary conditions
for a bar which is free at both ends, the analytical normal modes of example (b)
are identical to the elastic modes of a free bar.
\subsection{Two-dimensional segmentation}
\begin{figure}[h]
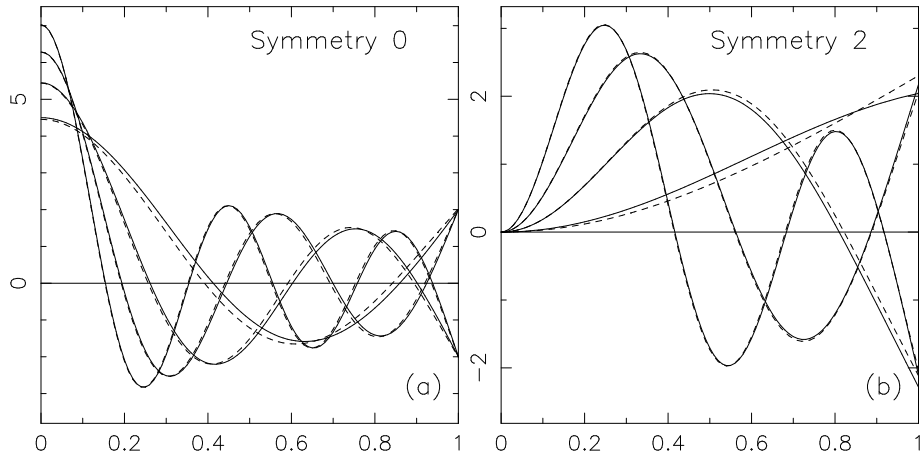

     \centerline{\hbox{
      \psfig{figure=figure13a.ps,width=60mm}
      \psfig{figure=figure13b.ps,width=60mm}}}
   \caption{\label{fig:figure13} {\small Comparison of the lowest analytical
     normal modes (solid lines) and corresponding elastic modes (dashed lines)
     for the rotational symmetries 0 and 2.}}
\end{figure}

It is shown in appendix A that the orthogonal set of minimum elastic energy modes of a
thin circular plate can be derived with a slightly more general method than the one
introduced in section~\ref{sec:introMethod}. 
For the set of elastic modes the corresponding set of curvature tensors has to be
orthogonal to the corresponding set of moment tensors.
This leads to the same differential equation (\ref{eq:PTTDiffEqFact})
as for the case of piston, tip and tilt movements in section~\ref{sec:derivModesPTT}.
Any differences in the shapes
must then be related to the boundary conditions.
For higher order modes with large wavenumbers
the terms with highest derivatives in the boundary conditions dominate the other terms.
If only the terms with the highest derivatives are retained, the boundary conditions
for the analytical normal modes in equation~(\ref{eq:boundaryFunctionalPTT})
and the elastic modes
in equations~(\ref{eq:bcEla1}) and (\ref{eq:bcEla2}) below of a free plate are identical :
\begin{eqnarray}
  \left( \frac{{\rm d}^{2}}{{\rm d}\rho^{2}}f_{i}(\rho) \right)_{\rho = 1} & = & 0
            \label{eq:bc1b}\\
  \left( \frac{{\rm d}^{3}}{{\rm d}\rho^{3}}f_{i}(\rho) \right)_{\rho = 1} & = & 0
            \label{eq:bc2b}
\end{eqnarray}
Therefore, within each rotational symmetry, the higher analytical normal modes should be
very similar to the corresponding elastic modes of a circular plate.
This similarity is obvious from figure~\ref{fig:figure13} which compares the
shapes of the lowest analytical normal modes of the rotational symmetries 0 and 2
with the shapes of the corresponding elastic modes.
\begin{center}
\begin{table}
\caption{ {\small Residual r.m.s. after fitting a set
of elastic modes of rotational symmetry 2 to the four lowest analytical normal modes of
rotational symmetry 2.}}
\begin{center}
\begin{tabular}{|lc||c|c|c|c|c|}
\hline
 & & \multicolumn{5}{|c|}{Fitted elastic modes} \\
         &  & 1 & 2 & 3 & 4 & 5 \\
\hline
       & 1 & 0.0817 & 0.0238 & 0.0104 & 0.0056 & 0.0023 \\
Normal & 2 & 0.9969 & 0.0053 & 0.0007 & 0.0007 & 0.0006 \\
modes  & 3 & 0.9998 & 0.9998 & 0.0124 & 0.0066 & 0.0042 \\
       & 4 & 1.0000 & 1.0000 & 0.9999 & 0.0156 & 0.0081 \\
\hline
\end{tabular}
\end{center}
\label{tab:table1}
\end{table}
\end{center}
Table~\ref{tab:table1} shows, for an initial r.m.s. of an analytical normal mode of 1,
the residual r.m.s. after fitting a given
number of the lowest elastic modes of a full circular plate to the lowest four
analytical normal modes
of rotational symmetry 2 of a full mirror.
The difference between the lowest analytical normal mode and the corresponding lowest elastic
mode is of the order of 8\% and one needs to fit the lowest three elastic modes to push
the residual r.m.s. to a level of 1\%. The similarity of the higher order modes is expressed by the
fact that the residual r.m.s. is effectively equal to 1 if the lowest $i-1$ elastic modes
are fitted to $i$-th analytical normal mode, but drops sharply to approximately 1\% after
the $i$-th elastic mode is included in the fit.

It is interesting to compare the probabilities of the occurence of normal modes
in a segmented mirror on the one hand with the probabilities of the occurence of elastic
modes in a monolithic mirror on the other hand. One has to choose the normal modes
obtained from the more general case of piston, tip and tilt movements presented in
section~\ref{sec:2DimAnalyticalPTT} because of their similarity to the
elastic minimum energy modes.

The source for wavefront errors in the form of
normal modes are the signals at intersegment edges. For white noise the r.m.s.
values $\epsilon_{{\rm sig},i}$ of all signal modes are identical. Since the
singular value is proportional to $1/\lambda_{i}^{2}$ the r.m.s.
$\epsilon_{{\rm nor},i}$ of a normal mode is related to $\epsilon_{\rm sig}$ by
\begin{equation}
  \epsilon_{{\rm nor},i} \propto \frac{\epsilon_{{\rm sig}}}{\lambda_{i}^{2}}
  \label{eq:relNS}
\end{equation}
The source for wavefront errors in the form of elastic modes are errors in the
support forces of the large monolithic mirrors. For a large number of supports these
force errors can be replaced by pressure fields.
Since the elastic modes form an orthogonal set of functions and
are proportional to the pressure fields generating them, the elastic modes and
pressure functions can be regarded as a pair of normal modes similar to the pair of
analytical signal and position modes in segmented mirrors. But, instead of being proportional
to $1/\lambda_{i}^{2}$ the r.m.s. $\epsilon_{{\rm ela},i} $ of an elastic mode is,
for a given r.m.s. $\epsilon_{{\rm pr}}$ of the pressure field, proportional to
$1/\lambda_{i}^{4}$ :
\begin{equation}
  \epsilon_{{\rm ela},i} \propto \frac{\epsilon_{{\rm pr}}}{\lambda_{i}^{4}}
  \label{eq:relEP}
\end{equation}
A comparison of the equations~(\ref{eq:relNS}) and (\ref{eq:relEP}) shows that the coefficients
of the elastic modes generated by random force errors decline much faster with the
order of the mode than the coefficients of the normal modes of a segmented mirror
generated by signal errors. The difference in the behaviour is, at a first glance,
surprising, since the normal modes of a segmented mirror are similar to the elastic modes,
that is the relationship between the position modes and the fields of the second
derivatives of the position modes is the same. But the difference is that for the normal
mirror modes the source is already a kind of curvature field, whereas for the elastic modes
there is another additional relationship involving a factor $\lambda_{i}^{2}$
between the source, which is the pressure field, and the curvature field.\\\\
{\large {\bf Appendix A : Derivation of elastic modes of a circular plate}}\\
For the derivation of the elastic modes of a free circular plate one has to generalise
the method introduced in section~\ref{sec:introMethod}.
Instead of equation~(\ref{eq:orthoLFLG})
\begin{equation}
           <Lf_{i}, Lf_{j}> \, \propto\,  \delta_{i,j}
	\label{eq:orthoLFLG1}
\end{equation}
we require now that
\begin{equation}
           <Lf_{i}, Nf_{j}> \, \propto\,  \delta_{i,j}
	\label{eq:orthoLFNG}
\end{equation}
with two possibly different differential operators $L$ and $N$. Partial integration
of the left hand side leads to
an equation which is equivalent to equation~(\ref{eq:integratedLfLf}) :
\begin{equation}
   <Lf_{i}, Nf_{j}> \, = \, <N^{F}Lf_{i}, f_{j}>
          + \, {\underline \beta}_{\rm L,N}[f_{i},f_{j}]
   \label{eq:integratedLfNf}
\end{equation}
If the boundary functional ${\underline \beta}_{\rm L,N}[f_{i},f_{j}]$
is zero, equations~(\ref{eq:orthoFG}), (\ref{eq:orthoLFNG}),
and (\ref{eq:integratedLfNf}) lead to the linear differential equation
\begin{equation}
   N^{F}Lf_{i} = \sigma^{2}_{i} \, f_{i}
     \label{eq:basicDiffEqLN}
\end{equation}
For elastic modes the operators are related to the strain and moment tensors.
With a prime and an asterix denoting a partial derivative with respect to the radial variable
$r$ and the azimuthal variable $\phi$, respectively,
the components of the strain tensor are defined by
 \begin {equation}
  \begin{array}{lll}
   {\displaystyle \chi_{{\rm r}}} = {\displaystyle-f''}, &
   {\displaystyle \chi_{\phi}} = {\displaystyle -\frac{f'}{r} - \frac{f^{**}}{r^{2}}}, &
   {\displaystyle\chi_{{\rm r}\phi}} =
             {\displaystyle-\frac{f^{'*}}{r} + \frac{f^{*}}{r^{2}}}
  \end{array}
 \end {equation}
and the components of the moment tensor by
 \begin{equation}
  \begin{array}{lll}
  M_{r} = D(\chi_{r} + \nu \chi_{\phi}), &
  M_{\phi} = D(\chi_{\phi} + \nu \chi_{r}), &
  M_{r\phi} = D(1-\nu)\chi_{r\phi} \; , \nonumber
 \end{array}
 \end{equation}
where $D = Kh^{2}/12$ and $K = Eh/(1-\nu^{2})$. $E$ is the
modulus of elasticity, $h$ the mirror thickness, and $\nu$ Poisson's ratio.
Equation~(\ref{eq:integratedLfNf}) then becomes
\begin{eqnarray}
   <Lf_{i}, Nf_{j}> & = & \int_{0}^{2\pi} {\rm d}\varphi
          \int_{r_{1}}^{r_{2}} {\rm d}r\,
        ( M_{r,i}\chi_{r,j} \, + \,
             M_{\phi,i}\chi_{\phi,j} \, + \,
             2M_{r\phi,i}\chi_{r\phi,j} ) \nonumber\\
       & = & \int_{r_{1}}^{r_{2}} {\rm d}r\, [
        (\chi_{r,i}\chi_{r,j} \, + \,
             \chi_{\phi,i}\chi_{\phi,j} \, + \,
             2\chi_{r\phi,i}\chi_{r\phi,j} )   \nonumber \\
       & &  \hspace*{20mm} + \, \nu \, (\chi_{\phi,i}\chi_{r,j} \, + \,
                     \chi_{r,i}\chi_{\phi,j} \, - \,
                     2\chi_{r\phi,i}\chi_{r\phi,j} )] \nonumber \\
       & \propto & \delta_{i,j}
\end{eqnarray}
Introducing equation~(\ref{eq:fTofn}) and partial integration lead to the differential
equation (\ref{eq:PTTDiffEqFact}) and a boundary functional.
For an arbitrary choice of the functions $f_{n,j}$ the boundary functional
is always zero if the following boundary conditions are fulfilled :
\begin{eqnarray}
   f''_{i}(r_{x}) \, + \, \nu \frac{f'_{i}(r_{x})}{r} \,
       - \, n^{2} \nu \frac{f_{i}(r_{x})}{r^{2}} & = & 0
    \label{eq:bcEla1}\\
   f'''_{i}(r_{x}) + \frac{f''_{i}(r_{x})}{r}
         + (1+n^{2}(2-v)) \frac{f'_{i}(r_{x})}{r^{2}}
              + n^{2}(3-v)\frac{f_{i}(r_{x})}{r^{3}} & = & 0
    \label{eq:bcEla2}
\end{eqnarray}
for both $r_{x} = r_{1}$ and $r_{x} = r_{2}$. These four boundary conditions are
identical to those for a free thin annular circular plate [3].\\\\
{\large {\bf Acknowledgements}} The author would like to thank Gary Chanan for many useful
discussions and Ray Wilson and Natalia Yaitskova for critical comments on the manuscript.\\\\

{\large {\bf References}}
\begin{itemize}
\item [1] Troy, M., Chanan, G., Sirko, E., and Leffert, E., 1998, Residual Misalignments of
the Keck Telescope Primary Mirror Segments : Classification of Modes and Implications for
Adaptive Optics, {\it Proceedings of the International Society for Optical Engineering},
 Vol. 3352, 307-317
\item [2] Noethe, L., Active optics in modern large optical telescopes,
       {\it Progress in Optics}, 43, chapter 1
\item [3] Noethe, L., 1991, Journal of Modern Optics, {\bf 38}, 1043-1066
\item [4] Cantrill, C.D., 2002, {\it Modern Mathematical Methods for Physicists
   and Engineers}, (Cambridge University Press)
\item [5] Chanan, G., 2002, Private communication
\item [6] Chanan, G., MacMartin, D.G., Nelson, J., Mast, T., Control and Alignment of
Segmented-Mirror Telescopes~: Matrices, Modes, and Error Propagation, {\it accepted for
 publication in Applied Optics}
\end{itemize}
\end{document}